\RequirePackage{fix-cm}
\documentclass[twoside, 11pt]{article}
\usepackage{geometry}
\usepackage{amsmath, amsthm, amssymb, dsfont, mathrsfs}
\usepackage{graphicx, caption, float, array, multirow, arydshln, chngcntr, etoolbox}
\usepackage{xcolor}
\usepackage{hyperref}
\hypersetup{colorlinks=true, linkcolor=blue, citecolor=blue, urlcolor=blue}

\counterwithin*{equation}{section}
\counterwithin*{equation}{subsection}

\usepackage{fourier, anyfontsize}
\usepackage[T1]{fontenc}

\allowdisplaybreaks

\usepackage{fancyhdr}
\newtheoremstyle{theorem}
  {10pt}
  {10pt}
  {\sl}
  {\parindent}
  {\bf}
  {. }
  { }
  {}
\theoremstyle{theorem}
\newtheorem{proposition}{Proposition}
\newtheorem{corollary}{Corollary}

\begin{document}

\title{\textbf{Power-type derivatives for rough volatility with jumps}}

\author{Liang Wang\thanks{Department of Mathematics and Statistics, Boston University Graduate School of Arts and Sciences. Email: \underline{leonwang@bu.edu}.} \and Weixuan Xia\thanks{Corresponding author; Department of Finance, Boston University Questrom School of Business. Email: \underline{gabxia@bu.edu}.}}
\date{2020}
\maketitle
\thispagestyle{plain}

\begin{abstract}
  In this paper we propose a novel pricing-hedging framework for volatility derivatives which simultaneously takes into account rough volatility and volatility jumps. Our model directly targets the instantaneous variance of a risky asset and consists of a generalized fractional Ornstein-Uhlenbeck process driven by a L\'{e}vy subordinator and an independent sinusoidal-composite L\'{e}vy process. The former component captures short-term dependence in the instantaneous volatility, while the latter is introduced expressly for rectifying the activity level of the average forward variance. Such a framework ensures that the characteristic function of average forward variance is obtainable in semi-closed form, without having to invoke any geometric-mean approximations. To analyze swaps and European-style options on average forward volatility, we introduce a general class of power-type derivatives on the average forward variance, which also provide flexible nonlinear leverage exposure. Pricing-hedging formulae are based on a modified numerical Fourier transform technique. A comparative empirical study is conducted on two independent recent data sets on VIX options, before and during the COVID-19 pandemic, to demonstrate that the proposed framework is highly amenable to efficient model calibration under various choices of kernels. \medskip\\
  MSC2020 Classifications: 60E10; 60G22; 60J76 \medskip\\
  JEL Classifications: C65; G13 \medskip\\
  \textsc{Key Words:} Rough volatility; volatility jumps; L\'{e}vy subordinators; sinusoidal processes; \\ power-type derivatives; VIX options
\end{abstract}

\newcommand{\dd}{{\rm d}}
\newcommand{\pd}{\partial}
\newcommand{\ii}{{\rm i}}
\newcommand{\E}{\mathbb{E}}
\newcommand{\PP}{\mathbb{P}}
\newcommand{\Var}{\mathrm{Var}}
\newcommand{\Skew}{\mathrm{skew}}
\newcommand{\EKurt}{\mathrm{ekurt}}
\newcommand{\Cov}{\mathrm{Cov}}
\newcommand{\EE}{\mathrm{E}}
\newcommand{\Gf}{\mathrm{\Gamma}}
\newcommand{\W}{\mathrm{W}}
\newcommand{\1}{\mathds{1}}
\newcommand{\U}{\varUpsilon}
\newcommand{\D}{\varDelta}
\renewcommand{\Sigma}{\varSigma}
\renewcommand{\Re}{\mathrm{Re}}
\renewcommand{\Im}{\mathrm{Im}}
\renewcommand{\Pi}{\varPi}

\section{Introduction}\label{sec:1}

``Rough volatility'' is a relatively new and yet already familiar jargon that has flourished in the financial world since the pioneering research work of [Gatheral et al, 2018] \cite{GJR}, which provided striking empirical evidence suggesting rough sample paths (compared to those of a semimartingale) of volatility observed in high-frequency financial time series and thus the presence of short-term dependence, while the idea of introducing frictions into volatility quantities goes back to the much earlier work of [Al\`{o}s et al, 2007] \cite{ALV} motivated from observations in option price-implied volatility surfaces. Over the past three years, a good number of works have been devoted to empirical justifications of rough volatility in various asset types. To name a few, [Livieri et al, 2018] \cite{LMPR} confirmed the existence of rough volatility by studying implied volatility-based approximations of spot volatility of the S\&P500 index, [Takaishi, 2020] \cite{T1} collected further evidence supporting volatility roughness in the cryptocurrency (in particular Bitcoin) market, and [Da Fonseca and Zhang, 2019] \cite{DZ} even demonstrated that rough volatility is also present in the VIX index.

Although the introduction of rough volatility has successfully reproduced stylized facts of historical volatility of asset prices, a series of difficulties have arisen in the meantime due to the loss of Markov and semimartingale properties. As a result, when developing pricing-hedging techniques accounting for rough volatility one will probably sojourn at Monte-Carlo simulation methods, whereas the inaccessibility of infinitesimal generators has disabled methods based on the Feynman-Kac formula. So far, simulation-based pricing-hedging methods have already been studied in depth; for example, [Jacquier et al, 2018] \cite{JMM} adopted a hybrid simulation scheme for the calibration of the rough Bergomi model initially proposed in [Bayer et al, 2016] \cite{BFG} on VIX futures and options. On the other hand, under the so-called ``rough Heston model'' which is constructed from a stationary power-type kernel and belongs to the family of affine Volterra processes discussed in [Jaber et al, 2019] \cite{JLP} (see also [Gatheral and Keller-Ressel, 2019] \cite{GKR}), characteristic function-based pricing methods were derived in [El Euch and Rosenbaum, 2019] \cite{ER2} which depend, partially, on solving a fractional Riccati equation, where their applicability was also demonstrated by a simple calibration exercise on S\&P500 implied volatility surfaces; the paper [El Euch and Rosenbaum, 2018] \cite{ER1} by the same authors considered from a theoretical standpoint similar hedging problems, after being able to write the characteristic function of the log-asset price in terms of a functional of its corresponding forward variance curve. We also notice the up-to-date work of [Horvath et al, 2020] \cite{HJT}, which adopted a martingale framework using forward variance curves in the goal of studying volatility options. It is worth mentioning that all these recent works have universally emphasized the role of a Brownian motion, having paid little attention to jumps in asset prices and their volatility, which are, of course, thought to complicate the pricing problems to great extent.\footnote{For instance, the aforementioned Riccati equation will turn into an integro-differential equation entailing more computationally expensive numerical schemes and the resultant model distributions will no longer be stable but subject to substantial changes under integral operations.}

All relevant models notwithstanding, one should however bear in mind that the key idea behind rough volatility is the exhibition of short-term dependence, or more precisely, rapidly decaying autocorrelation near the origin, rather than inherent reliance on Brownian sample paths, or path continuity, which characteristic is arguably an estimation assumption imposed in [Gatheral et al, 2018] \cite{GJR} and deemed nonessential. In fact, extensive use of the Brownian motion in the cited literature is more or less an act of simplicity, mainly due to log-instantaneous volatility shown to be empirically close to normally distributed. On the other hand, disregarding the exclusive use of the Brownian motion sheds light upon another important aspect -- the presence of volatility jumps. In a semimartingale setting, this would send us back to the work of [Todorov and Tauchen, 2011] \cite{TT}, which, by analyzing from high-frequency VIX index data the activity level of some presumed mean-reverting instantaneous variance model, showed that stock market volatility should be most suitably depicted as a purely discontinuous process without a Brownian component. Notably, this concern may seem inconsequential in a non-semimartingale model with frictions, as pointed out in the same paper: In short, the activity level of the process can be flexibly adjusted according to the controlling fraction parameter. For this reason, inclusion of volatility jumps in a model that is already fractional has seemingly been ignored for investigation. Nevertheless, since increased activity levels are an inevitable consequence of increased path roughness, using a fractional Brownian motion with a fraction parameter less than 1 will only increase the activity level of the resultant variance process, which to a degree neglects the empirical findings of [Todorov and Tauchen, 2011] \cite{TT}; see also [Bollerslev and Todorov, 2011] \cite{BT} and [Bardgett et al, 2019] \cite{BGL} for similar confirmations of the necessity of volatility jumps. In connection with this, we expect that replacing the Brownian motion with a purely discontinuous process whose sample paths are less active, combined with suitable modifications, is able to strike a balance between these two important aspects (short-term dependence and volatility jumps) and eventually yield desirable modeling outcomes.

These inspire us to take on a new path deviating from the use of a fractional Brownian motion in the establishment of rough volatility and switch to purely discontinuous square-integrable L\'{e}vy processes of infinite activity. In more detail, we want to propose a flexible framework for the instantaneous variance based on the sum of a generalized fractional Ornstein-Uhlenbeck process subject to an integrable kernel and an independent bounded process; a key feature of this formulation is that it is not derived from taking logarithms but yet is capable of simultaneously capturing short-term dependence and possible jumps in the instantaneous variance, as well as achieving an arbitrary suitable activity level of the corresponding forward variance curve. It is fundamentally a quasi-linear framework resembling the well-known BNS models of [Barndorff-Nielsen and Shephard, 2001] \cite{B-NS}. We note that models of non-exponential type have also been widely applied in volatility analysis, some recent developments including [Hofmann and Schulz, 2016] \cite{HS} and [Issaka and SenGupta, 2017] \cite{IS}. Needless to say, despite that inclusion of jumps may not result in significant improvement of the model fit of volatility distributions observed at high frequencies, as noted in [Gatheral et al, 2018, \text{Sect.} 6] \cite{GJR}, it is undoubtedly innocuous and the model distribution in logarithm can also become arbitrarily close to normality by properly tuning scale parameters, thanks to the central limit theorem. On the contrary, introducing jumps into the instantaneous variance gives rise to an analytically tractable structure for the characteristic function of the average forward volatility, facilitating the pricing and hedging of volatility derivatives of interest. In particular, such a structure requires no inexact transformations, such as the geometric-mean approximation adopted in \text{e.g.} [Horvath et al, 2020] \cite{HJT}, for the average forward volatility, which appear to be inevitable under exponential models.

Besides, although our main results are given in a general setting, attention will be drawn to three particular types of stationary kernels, all of which are comfortable to work with and have their own advantages. While the first type is recognized for its incommensurable simplicity, the second is compatible with the transformation of the instantaneous variance dynamics into a usual Ornstein-Uhlenbeck process but driven by a fractional L\'{e}vy process. The third type is arguably a result of reverse engineering and designed specifically to avoid certain transcendental functions that are relatively costly to implement. In so doing it will be interesting as well to compare the overall suitability of various types of kernels, despite their considerable similarity to each other in shape.

As already noted, our ultimate objective in the present paper lies in analyzing European-style financial derivatives written on the average forward volatility, such as the VIX index, including swaps and options. Under the proposed model framework, we obtain pricing-hedging formulae for a more general class of power-type derivatives, which raise the underlying volatility or the standard option payoff to a certain nonnegative power. Noteworthily, derivatives with power payoff functions written on equity have been thoroughly examined in the literature; see [Tompkins, 1999] \cite{T2}, [Raible, 2000] \cite{R1}, [Macovschi and Quittard-Pinon, 2006] \cite{MQ-P}, and [Xia, 2017] \cite{X1} on single-asset options and [Blenman and Clark, 2005] \cite{BC}, [Wang, 2016] \cite{W}, and [Xia, 2019] \cite{X2} on exchange options. Similar exchange options on zero-coupon bonds have recently been studied in [Blenman et al, 2020] \cite{BB-GC}. Along these lines, consideration of power-type derivatives in the volatility market also has significance in generating nonlinear leverage effects on the investor's risk exposure, and will be conducive to hedging volatility-of-volatility risks. Implementation of the proposed pricing-hedging formulae will heavily rely on numerical Fourier transform techniques.

The remainder of this paper is organized as follows. In Section \ref{sec:2} we establish our model framework starting from the instantaneous variance dynamics and provide a comprehensive analysis of its properties, including covariance function and path regularity, and then give an integral representation for the characteristic function of the average forward variance. Some simulation techniques are discussed in Section \ref{sec:3}, with pertinent convergence results. Section \ref{sec:4} contains our new pricing-hedging formulae for power-type derivatives that nonlinearly extend standard volatility derivatives, which then initiate a comparative empirical study in Section \ref{sec:5} focused around VIX options, utilizing two independent data sets and two of the proposed kernels. Last but not least, we also provide some insight into how the model framework may be further extended to accommodate the presence of rough volatility of volatility in Section \ref{sec:6} by means of a stochastic time change argument, in catering for the noted finding of [Da Fonseca and Zhang, 2019] \cite{DZ}. Conclusions and future research directions are outlined in Section \ref{sec:7} and all mathematical proofs presented in the end.

\section{Construction of rough volatility with jumps}\label{sec:2}

\subsection{Fractional L\'{e}vy processes}\label{sec:2.1}

We begin by synthesizing some crucial ingredients of a non-Gaussian fractional L\'{e}vy process, which are necessary for establishing a model for the instantaneous variance of a risky asset whose sample paths have roughness and jump features. Of course, allowing for positivity of the instantaneous variance the background-driving L\'{e}vy process must be nonnegative, i.e., a subordinator.

To this end, consider a continuous-time stochastic basis $\mathfrak{S}:=(\Omega,\mathcal{F},\PP;\mathbb{F}\equiv\{\mathscr{F}_{t}\}_{t\geq0})$, where the filtration $\mathbb{F}$ is assumed to satisfy the usual conditions. Let $X\equiv(X_{t})$ be an adapted and square-integrable L\'{e}vy subordinator supported on $\mathfrak{S}$, which is exclusively characterized by a Poisson random measure $N_{X}$ defined on $(\mathds{R}_{++},\mathds{R}_{+})$. According to the L\'{e}vy-Khintchine representation, $X_{1}$ has the characteristic exponent
\begin{equation*}
  \log\phi_{X_{1}}(l):=\log\E\big[e^{\ii lX_{1}}\big]=\int^{\infty}_{0+}(e^{\ii lz}-1)\nu_{X}(\dd z),\quad l\in\mathds{R},
\end{equation*}
where $\ii$ denotes the imaginary unit and $\nu_{X}$ is the intensity measure associated with $N_{X}$. For practicality we impose the assumption that $\nu$ is non-atomic so that the distribution of $X_{1}$ is absolutely continuous with respect to Lebesgue measure (see, e.g., [Kohatsu-Higa and Takeuchi, 2019, Theorem 6.3.4] \cite{K-HT}) and we denote by $\xi_{1}:=\E[X_{1}]>0$ and $\xi_{2}:=\Var[X_{1}]>0$. Since $X$ has independent and stationary increments, it has the familiar covariance function, for any $u>0$,  $\Cov\big[X_{t},X_{t+u}\big]=\xi_{2}t$.

For a continuously differentiable\footnote{Continuous differentiability is a highly desirable property of the kernel for modeling purposes, hence assumed throughout this paper. Intuitively, by ruling out kinks and discontinuities it ensures that the frictions brought by $g$ do not have sudden changes. However, it is not required for defining fractional L\'{e}vy subordinators and thus not to be comprehended as any implicit assumption for the ongoing analysis.} kernel $g\in\mathcal{C}^{(1,1)}$ defined in the domain $\{(t,s):t>0,s\in[0,t)\}$ satisfying the integrability condition
\begin{equation*}
  \int^{t}_{0}g^{2}(t,s)\dd s<\infty,\quad\forall t>0,
\end{equation*}
we then define the fractional L\'{e}vy process via the following Volterra-type stochastic integral,
\begin{equation}\label{2.1.1}
  X^{(g)}_{t}:=\int^{t}_{0}g(t,s)\dd X_{s}=\int^{t}_{0}\int^{\infty}_{0+}g(t,s)zN_{X}(\dd z,\dd s),\quad t\geq0,
\end{equation}
which is well-defined $\PP$-a.s., and is \text{a.k.a.} a L\'{e}vy-driven fractionally integrated moving-average process (see [Marquardt, 2006, \text{Sect.} 6.2] \cite{M1}). With this representation, for any $t,u>0$, it is easy to see that $\E\big[X^{(g)}_{t}\big]=\xi_{1}\int^{t}_{0}g(t,s)\dd s$ and, by using the L\'{e}vy-It\^{o} isometry (see, e.g., [Lyasoff, 2017, \text{Sect.} 16.32] \cite{L2}), the covariance function of the process $X^{(g)}$ takes the following form,
\begin{equation*}
  \Cov\big[X^{(g)}_{t},X^{(g)}_{t+u}\big]=\xi_{2}\int^{t}_{0}g(t,s)g(t+u,s)\dd s.
\end{equation*}
Notably, if $\lim_{s\nearrow t}g(t,s)=\infty$ for every $t>0$, then the integral $\int^{t}_{0}g^{(0,1)}(t,s)\dd s$ is divergent for every $t>0$. In this case, $X^{(g)}$ exhibits short-term dependence in the sense that there exists $\varpi\in(0,1)$ such that
\begin{equation*}
  \Cov\big[X^{(g)}_{t},X^{(g)}_{t+u}\big]=\Var\big[X^{(g)}_{t}\big]+\xi_{2}C(t)u^{\varpi}+O(u),
\end{equation*}
where $\Var\big[X^{(g)}_{t}\big]=\xi_{2}\int^{t}_{0}g^{2}(t,s)\dd s$ by the dominated convergence theorem and $C(t)$ is some constant depending only on $t>0$.

To give a few examples, the Molchan-Golosov kernel ([Molchan and Golosov, 1969] \cite{MG}) reads
\begin{equation}\label{2.1.2}
  g(t,s)=(t-s)^{d-1}\;_{2}\mathrm{F}_{1}\bigg(-d,d-1;d;-\frac{t-s}{s}\bigg),
\end{equation}
for some fraction parameter $d\in(1/2,3/2)$, where $\;_{2}\mathrm{F}_{1}(\cdot,\cdot;\cdot;\cdot)$ is the Gauss hypergeometric-$(2,1)$ function ([Abramowitz and Stegun, 1972, \text{Sect.} 15] \cite{AS}) and which is non-stationary. We note that the specific form (\ref{2.1.2}) was initially chosen in [Jost, 2006] \cite{J} in an attempt to match the Weyl integral representation of a fractional Brownian motion living in real-valued time used in [Mandelbrot and van Ness, 1968] \cite{MvN}. It was shown in [Tikanm\"{a}ki and Mishura, 2011] \cite{TM}, however, that such transformation does not necessarily lead to the same finite-dimensional distribution in the more general case of fractional L\'{e}vy processes. Another popular choice of $g$ is the following obviously stationary and yet structurally much simpler Riemann-Liouville kernel,\footnote{The same kernel was used in [El Euch, 2018] \cite{ER1} and [El Euch, 2019] \cite{ER2} in constructing the (generalized) rough Heston model.}
\begin{equation}\label{2.1.3}
  g(t,s)\equiv g(t-s)=\frac{(t-s)^{d-1}}{\Gf(d)},
\end{equation}
for $d>1/2$, where $\Gf(\cdot)$ denotes the usual gamma function. In particular, with (\ref{2.1.3}) the fractional process $X^{(g)}$ can also be understood as a consequence of repeated path integration of the subordinator $X$, i.e., for $d\in\mathds{N}_{++}\equiv\mathds{N}\setminus\{0\}$,
\begin{equation}\label{2.1.4}
  X^{(g)}_{t}=\underbrace{\int\cdots\int^{t}_{0}}_{d}X_{s}\underbrace{\dd s\dots\dd s}_{d},
\end{equation}
which can be extended through Cauchy's repeated integration formula. Of course, for this special choice roughness is only present if $d\in(1/2,1)$, while it also has the obvious drawback, compared to the Molchan-Golosov kernel, that the resultant fractional process fails to have stationary increments.

In any case, a well-suited candidate for $X$, having infinitely many jumps on compact time intervals, can be a one-sided tempered stable process, i.e., a tempered stable subordinator, which has three parameters -- $a>0$, $b>0$ and $c\in(0,1)$, leading to the following characteristic exponent,
\begin{equation}\label{2.1.5}
  \log\phi_{X_{1}}(l)=a\Gf(-c)((b-\ii l)^{c}-b^{c}),\quad l\in\mathds{R},
\end{equation}
so that $\nu(\dd z)=ae^{-bz}/z^{c+1}\1_{(0,\infty)}(z)\dd z$, for $z>0$, which is clearly an infinite measure. The tempered stable distribution constitutes a fairly general family of infinitely divisible distributions (see [Rosi\'{n}ski, 2007] \cite{R2} and [K\"{u}chler and Tappe, 2013] \cite{KT}, as well as the overview in [Schoutens, 2003, \text{Sect.} 5.3] \cite{S1}), and it has witnessed many applications in constructing volatility models in discrete time (see, e.g., [Mercuri, 2008] \cite{M2} and [Li et al, 2016] \cite{LLZ}). In particular, by taking $c\searrow0$ and $c=1/2$, from $X$ one recovers the well-known gamma process and inverse Gaussian process, respectively; in the former case it is understood that $\log\phi_{X_{1}}(l)=-a\log(1-\ii l/b)$. With (\ref{2.1.5}) it is also straightforward to verify that $\xi_{1}=a\Gf(1-c)/b^{1-c}$ and $\xi_{2}=a\Gf(2-c)/b^{2-c}$.

\subsection{Instantaneous variance}\label{sec:2.2}

Let us consider, instead of the natural logarithm of the instantaneous volatility of a risky asset, the instantaneous variance process, denoted $V\equiv(V_{t})$. Intuitively speaking, our idea is to express $V$ as a Volterra-type stochastic integral, up to shifting and positive scaling, analogous to the fractional L\'{e}vy process $X^{(g)}$ in (\ref{2.1.1}) with a suitable kernel chosen to allow for short- or long-term dependence as well as long-term mean reversion, which is then supplemented by an independent bounded semimartingale to adjust the vibrancy of the sample paths. This is done by assuming the following quasi-Ornstein-Uhlenbeck structure,
\begin{equation}\label{2.2.1}
  V^{\circ}_{t}=V^{\circ}_{0}e^{-\kappa t}+\bar{V}(1-e^{-\kappa t})+X^{(h)}_{t},\quad V_{t}=V^{\circ}_{t}+\varsigma(\cos Z_{t}+1),\quad t\geq0,
\end{equation}
whose ingredients are set up as follows: $\kappa>0$ specifies a reversion speed and $\bar{V}\geq0$ a universal reversion level, $h$ is a continuously differentiable kernel having a power-law left tail and a power-exponential right tail,
\begin{equation}\label{2.2.2}
  h(t+u,t)=
  \begin{cases}
    O(u^{d-1}),\quad\text{as }u\searrow0,\\
    O\big(e^{-\kappa u}u^{(d-1)^{+}}\big),\quad\text{as }u\rightarrow\infty,
  \end{cases}
  \quad\forall t>0,
\end{equation}
for a fraction parameter $d>1/2$, where $(\cdot)^{+}$ denotes the positive part; $Z$ is a real-valued purely discontinuous L\'{e}vy process independent from $X$, also supported on $\mathfrak{S}$, with an absolutely continuous distribution for every fixed $t>0$, and $\varsigma>0$ is a small (manually adjustable) scaling factor. Subject to the law continuity assumption it suffices that $Z_{1}$ admit a characteristic exponent of the form
\begin{equation}\label{2.2.3}
  \log\phi_{Z_{1}}(l):=\log\E\big[e^{\ii lZ_{1}}\big]=\int_{\mathds{R}\setminus\{0\}}(e^{\ii lz}-1-\ii lz\1_{\{|z|\leq 1\}})\nu_{Z}(\dd z),\quad l\in\mathds{R},
\end{equation}
whenever the above integral is finite, where $\nu_{Z}$ is the L\'{e}vy measure of $Z$ with $\int_{0<|z|<1}|z|\nu_{Z}(z)=\infty$.

The condition (\ref{2.2.2}) subtly embodies the intuition of introducing frictions into $V$ without jeopardizing its mean-reverting property, and it automatically ensures that (\ref{2.2.1}) is well defined because $\sup_{t>0}\int^{t}_{0}h^{2}(t,s)\dd s<\infty$. Besides, we assume that $V^{\circ}_{0}>0$ has a known value. In fact, if (\ref{2.2.2}) holds with $d\in(1/2,1)$, then $V$ exhibits mean reversion in the long term but is simultaneously allowed to have short-term dependence. As before, with the L\'{e}vy-It\^{o} isometry the process $V$ is seen to have the covariance function equal to
\begin{equation*}
  \Cov[V_{t},V_{t+u}]=\xi_{2}\int^{t}_{0}h(t,s)h(t+u,s)\dd s+\varsigma^{2}\Cov[\cos Z_{t},\cos Z_{t+u}],
\end{equation*}
for any $u>0$, which generally depends on $t>0$. Under (\ref{2.2.2}), if $d<1$ then $\lim_{s\nearrow t}h(t,s)=\infty$ for any $t>0$ and there exists $C(t)\in\mathds{R}$ depending only on $t$ such that
\begin{equation}\label{2.2.4}
  \Cov[V_{t},V_{t+u}]=\Var[V_{t}]+\xi_{2}C(t)u^{2d-1}+O(u),\quad\text{as }u\searrow0,
\end{equation}
with $\Var[V_{t}]=\xi_{2}\int^{t}_{0}h^{2}(t,s)\dd s+\varsigma^{2}\Var[\cos Z_{t}]$. In this case, the covariance function is rough at the origin and $V$ exhibits short-term dependence. With the right-tail behavior in (\ref{2.2.2}) $V$ always reverts to a positive mean in the long term, with
\begin{equation}\label{2.2.5}
  \lim_{t\rightarrow\infty}\E[V_{t}]=\bar{V}+\xi_{1}\lim_{t\rightarrow\infty}\int^{t}_{0}h(t,s)\dd s+\varsigma>0,
\end{equation}
where we have used $\lim_{t\rightarrow\infty}\E[\cos Z_{t}]=0$ as $\big|\E\big[e^{\ii Z_{1}}\big]\big|<1$ due to the absolute continuity of the distribution of $Z_{1}$.

Observably, our construction of the instantaneous variance $V$ in (\ref{2.2.1}) consists of two components: a generalized fractional Ornstein-Uhlenbeck process $V^{\circ}$ and a composite-sinusoidal L\'{e}vy process $\varsigma(\cos Z+1)$. Driven by a subordinator ($X$), the first component accounts for large upward volatility jumps but depicts significant downward movements as continuous corrections. This is notably in line with recent empirical evidence (see, e.g., [Park, 2016] \cite{P1}) about the relative importance of signed jumps. Meanwhile, the latter component is included for the purpose of capturing small-scale two-sided volatility movements of suitable activity. Simplicity aside, choosing the cosine function has a particular advantage in that it does not alter the nature of (infinite-variation) small jumps of $Z$, albeit wiping out large values, for, with the L\'{e}vy-It\^{o} decomposition in mind, $\cos'z=-\sin z=O(z)$ as $z\rightarrow0$.

The fractional part of the process $V^{\circ}$ is actually motivated by the recipe used in [Wolpert and Taqqu, 2004, \text{Sect.} 3] \cite{WT} based on repeated integration, where $h$ is specialized as the product of a usual exponential kernel and the Riemann-Liouville kernel,
\begin{equation}\label{2.2.6}
  h(t,s)\equiv h(t-s)=\frac{e^{-\kappa(t-s)}(t-s)^{d-1}}{\Gf(d)},
\end{equation}
which is obviously strictly positive and will be referred to as the type-I kernel. A remarkable difference, however, is that we assume that the instantaneous variance process is only observed starting from time 0.

Indeed, the structure (\ref{2.2.1}) represents a wide range of approaches towards achieving path roughness and mean reversion at the same time, while it gives rise to an Ornstein-Uhlenbeck process driven by a fractional L\'{e}vy process (subordinator), i.e., the structure used in [Garnier and S{\o}lna, 2018] \cite{GS}, with the choice
\begin{equation}\label{2.2.7}
  h(t,s)=g(t,s)-\kappa\int^{t}_{s}e^{-\kappa(t-v)}g(v,s)\dd v,
\end{equation}
where $g$ is the kernel mentioned in (\ref{2.1.1}). With an application of It\^{o}'s formula and the Fubini-Tonelli theorem the first equation in (\ref{2.2.1}) is reformatted into
\begin{equation}\label{2.2.8}
  V^{\circ}_{t}=V^{\circ}_{0}e^{-\kappa t}+\bar{V}(1-e^{-\kappa t})+\int^{t}_{0}e^{-\kappa(t-s)}\dd X^{(g)}_{s},
\end{equation}
which is the solution of the fractional stochastic integral equation
\begin{equation*}
  V^{\circ}_{t}=\kappa\int^{t}_{0}(\bar{V}-V^{\circ}_{s})\dd s+X^{(g)}_{t}.
\end{equation*}
Since the second term on the left-hand side of (\ref{2.2.7}) is bounded for every fixed $t>0$, it holds that $\lim_{s\nearrow t}h(t,s)/g(t,s)>0$, provided $\lim_{s\nearrow t}g(t,s)=\infty$ for any $t>0$. Thus, if the sample paths of $X^{(g)}$ exhibit short-term dependence, then so do those of $V$, and in fact, their short-term dependence must be of the same degree. If $g$ is further taken to be the stationary Riemann-Liouville kernel, then by straightforward calculations (\ref{2.2.7}) yields the following type-II kernel which also happens to be stationary,
\begin{equation}\label{2.2.9}
  h(t,s)\equiv h(t-s)=\frac{(t-s)^{d-1}+(-\kappa)^{1-d}e^{-\kappa(t-s)}(\Gf(d)-\Gf(d,-\kappa(t-s)))}{\Gf(d)},
\end{equation}
where $\Gf(\cdot,\cdot)$ denotes the upper incomplete gamma function. The correlation structure (\ref{2.2.4}) can be made more precise with $\Var\big[X^{(h)}_{t}\big]=\xi_{2}\int^{t}_{0}h^{2}(s)\dd s$ and some
\begin{equation}\label{2.2.10}
  C(t)\in\frac{\Gf(1-2d)\sin(\pi d)}{\pi\Gf^{2}(d)}\times(e^{-2\kappa t},1),
\end{equation}
which only depends on $t$. This shows that roughness is established if and only if $d\in(1/2,1)$. However, formed from the Riemann-Liouville kernel the type-II kernel fails to meet the right tail behavior condition imposed in (\ref{2.2.2}), but obeys instead $h(t+u,t)\equiv h(u)=O\big(e^{-\kappa\1_{\{d=1\}}u}u^{(d-2)\1_{\{d\neq1\}}}\big)$ as $u\rightarrow\infty$, $\forall t>0$. This signifies that the Ornstein-Uhlenbeck process $V^{\circ}$ driven by a fractional L\'{e}vy subordinator based on the Riemann-Liouville kernel can be mean-reverting only if $d\leq1$, which impedes long-term dependence. Hence, it is no surprise that with this kernel the long-term mean (\ref{2.2.4}) of the instantaneous variance is finite only for $d\in(1/2,1]$, and is simply given by $\lim_{t\rightarrow\infty}\E[V_{t}]=\bar{V}+\1_{\{d=1\}}/\kappa+\varsigma$ (as $\int^{\infty}_{0}h(s)\dd s=\1_{\{d=1\}}/\kappa$), thereby indicating that the type-II kernel, unlike the type-I, is not strictly positive when short-term dependence is exhibited with $d<1$, but the resultant process $V$ obviously is (by (\ref{2.2.8})).

With the type-I kernel (\ref{2.2.6}), it is not possible to interpret $V^{\circ}$ as an Ornstein-Uhlenbeck process driven by a fractional L\'{e}vy process. Nonetheless, $V$ can still have short-term dependence since, for $d\in(1/2,1)$,
\begin{equation*}
  \frac{\Cov\big[V_{t},V_{t+u}\big]}{\xi_{2}}=\frac{t^{2d-1}}{(2d-1)\Gf^{2}(d)}+C(t)u^{2d-1}+O(u),\quad\text{as }u\searrow0,
\end{equation*}
by using (\ref{2.2.10}). The long-term mean is however unconditionally finite: $\lim_{t\rightarrow\infty}\E[V_{t}]=\bar{V}+\xi_{1}\kappa^{-d}+\varsigma$.

As another aspect of our innovation, for $d\in(1/2,1)$ generating short-term dependence we propose to construct $h$ by combining the scaled exponential kernel and the Riemann-Liouville kernel in a piecewise fashion, i.e.,
\begin{equation}\label{2.2.11}
  h(t,s)\equiv h(t-s)=
  \begin{cases}
    \displaystyle \frac{(t-s)^{d-1}-\tau^{d-1}}{\Gf(d)}+\theta e^{-\kappa\tau}&\quad\text{if }t-s<\tau,\\
    \displaystyle \theta e^{-\kappa(t-s)}&\quad\text{if }t-s\geq\tau,
  \end{cases}
\end{equation}
where $\tau>0$ is some time threshold separating the power-law and exponential parts of the kernel and $\theta>0$ is some scaling factor. To ensure continuous differentiability, $\tau$ solves the transcendental equation
\begin{equation}\label{2.2.12}
  e^{\kappa\tau}\tau^{d-2}=-\kappa\theta\Gf(d-1).
\end{equation}
In order for (\ref{2.2.12}) to be solvable on $\mathds{R}_{++}$, $\theta$ has to satisfy the constraint
\begin{equation}\label{2.2.13}
  \theta\geq-\frac{1}{\kappa\Gf(d-1)}\bigg(\frac{2-d}{e\kappa}\bigg)^{d-2},
\end{equation}
under which
\begin{equation}\label{2.2.14}
  \tau=\frac{d-2}{\kappa}\W_{i}\bigg(\frac{\kappa}{d-2}(-\kappa\theta\Gf(d-1))^{1/(d-2)}\bigg),\quad i\in\{-1,0\},
\end{equation}
with $\W_{\cdot}(\cdot)$ being the Lambert W function, \text{a.k.a.} the product logarithm (see [Corless et al, 1996] \cite{CGHJK}). Note that the two solutions in (\ref{2.2.14}) coincide if and only if equality holds in (\ref{2.2.13}). We will refer to (\ref{2.2.11}) as the type-III kernel.\footnote{Such a piecewise construction has the same shortcoming as the type-II kernel, namely the restriction of $d\in(1/2,1)$, but instead of sacrificing the mean-reverting property, the type-III kernel is incompatible with differentiability at $\tau$ if long-term dependence ($d\geq1$) is required, in which case (\ref{2.2.12}) has no positive solutions.} Before the type-III kernel can be properly compared to the other two types, we note that the type-I and type-II kernels both have the parametrical limiting property that $\lim_{\kappa\searrow0}h(t,s)=(t-s)^{d-1}/\Gf(d)$ and $\lim_{d\nearrow1}h(t,s)=e^{-\kappa(t-s)}$, which correspond to the Riemann-Liouville kernel without mean reversion and the classical exponential kernel without fractions, respectively. It is desirable that the same hold for the type-III kernel, which is possible by its construction. The particular choice of $(\tau,\theta)$ that achieves this effect is also not difficult to find by the properties of the Lambert W function, and is given by
\begin{equation*}
  \tau=\frac{1-d}{\kappa}\quad\text{and}\quad\theta=-\frac{(e\kappa)^{1-d}}{(1-d)^{2-d}\Gf(d-1)}.
\end{equation*}
Then the type-III kernel is uniquely parameterized by $\kappa$ and $d$ and reads
\begin{equation}\label{2.2.15}
  h(t-s)=
  \begin{cases}
    \displaystyle \frac{(t-s)^{d-1}-((1-d)/\kappa)^{d-1}}{\Gf(d)}-\frac{\kappa^{1-d}}{(1-d)^{2-d}\Gf(d-1)}&\quad\displaystyle \text{if }t-s<\frac{1-d}{\kappa},\\
    \displaystyle -\frac{(e\kappa)^{1-d}e^{-\kappa(t-s)}}{(1-d)^{2-d}\Gf(d-1)}&\quad\displaystyle \text{if }t-s\geq\frac{1-d}{\kappa}.
  \end{cases}
\end{equation}
With (\ref{2.2.15}), we have in (\ref{2.2.4}) that $C(t)\equiv C=\Gf(1-2d)\sin(\pi d)/(\pi\Gf^{2}(d))$ while the long-term mean (\ref{2.2.4}) is always finite:
\begin{equation*}
  \lim_{t\rightarrow\infty}\E[V_{t}]=\bar{V}+\frac{1}{(1-d)\Gf(d+1)}\bigg(\frac{1-d}{\kappa}\bigg)^{d}+\varsigma>0.
\end{equation*}

\begin{figure}[H]
  \centering
  \includegraphics[width=2.9in]{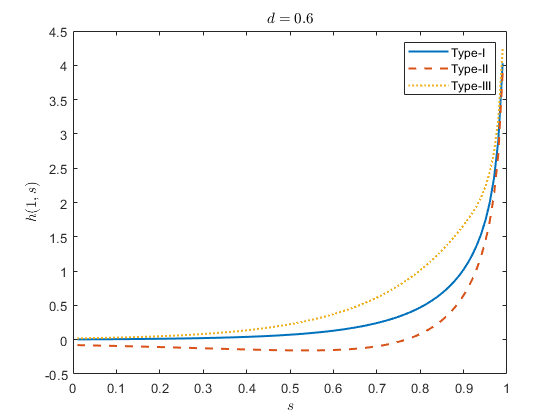}
  \includegraphics[width=2.9in]{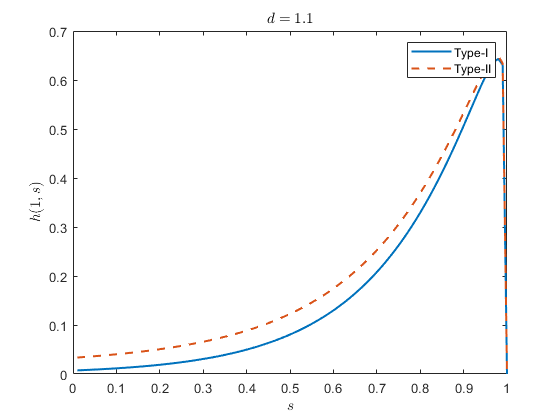}
  \caption{Comparison of kernels}
  \label{fig:1}
\end{figure}

Using $\kappa=5$ and $d=0.6$, Figure \ref{fig:1} below compares the three types of kernels (\ref{2.2.6}), (\ref{2.2.9}) and (\ref{2.2.15}) over the unit time interval. It is clear that they share the same right-tail behavior, generating the same degree of short-term dependence. In fact, with $d<1$ the type-I and type-III kernels are both monotone and strictly positive whereas the type-II kernel is not.

On a different note, despite that by the formulation (\ref{2.2.1}) $V$ lacks increment stationarity, there is comprehensibly no negative impact placed on characteristic function-based model calibration.

In the next proposition we give some partial results on the regularity of the sample paths of $V$. Indeed, from its appellation the interpretation of path roughness is not confined to the involvement of short-term dependence, but should be linked to how degree of irregularity of the sample paths. Since $\cos Z$ is a purely discontinuous semimartingale, we focus on the generalized fractional Ornstein-Uhlenbeck process $V^{\circ}$ and consider the case of an infinite L\'{e}vy measure $\nu_{X}$ allowing for practicality.

\begin{proposition}\label{prop:1}
Assume $\nu_{X}(\mathds{R}_{++})=\infty$ and (\ref{2.2.1}). For any fixed time $T>0$ we have the following three assertions.

(i) If $d>1$, then the sample paths of $V^{\circ}$ are $\PP$-\text{a.s.} continuous with $\PP$-\text{a.s.} zero quadratic variation over $[0,T]$.

(ii) If $d=1$, then the sample paths of $V^{\circ}$ are $\PP$-\text{a.s.} discontinuous with $\PP$-\text{a.s.} finite quadratic variation over $[0,T]$.

(iii) If $1/2<d<1$, then the sample paths of $V^{\circ}$ are $\PP$-\text{a.s.} discontinuous and unbounded with $\PP$-\text{a.s.} infinite quadratic variation over $[0,T]$.
\end{proposition}

Notably, for $d>1$, the sample paths of $V^{\circ}$ are smoothed in a way that all the jumps generated by $X$ are expunged and, as will be seen in the proof in Appendix A, they are actually H\"{o}lder-continuous for suitable exponents. On the other hand, in the situation of assertion (iii), $V$ can have infinitely large jumps, so that its sample paths form maps from $[0,T]$ to $[0,\infty]$. However, this will not be a problem for modeling in practice because $V_{t}$ is \text{a.s.} finite for any fixed $t\geq0$ and in fact, it has a finite variance. For the critical value $d=1$, $V$ only exhibits mean reversion. For the type-I and type-II kernels it can be easily verified that in the case $d=1$ $V$ is exactly the usual L\'{e}vy-driven Ornstein-Uhlenbeck process and for the type-III kernel this is also true in the limit as $d\nearrow1$.

Letting $t_{0}\in[0,t]$ be a fixed time point, we can recast (\ref{2.2.1}) conditional on $\mathscr{F}_{t_{0}}$ as
\begin{align}\label{2.2.16}
  V_{t}&=V^{\circ}_{0}e^{-\kappa t}+\bar{V}(1-e^{-\kappa t})+\int^{t_{0}}_{0}h(t,s)\dd X_{s}+\int^{t}_{t_{0}}h(t,s)\dd X_{s} \nonumber\\
  &\qquad+\varsigma(\cos(Z_{t}-Z_{t_{0}})\cos Z_{t_{0}}-\sin(Z_{t}-Z_{t_{0}})\sin Z_{t_{0}}+1),
\end{align}
or equivalently,
\begin{align}\label{2.2.17}
  V_{t}&=V^{\circ}_{t_{0}}e^{-\kappa(t-t_{0})}+\bar{V}(1-e^{-\kappa(t-t_{0})})+\int^{t_{0}}_{0}\big(h(t,s)-e^{-\kappa(t-t_{0})}h(t_{0},s)\big)\dd X_{s}+\int^{t}_{t_{0}}h(t,s)\dd X_{s} \nonumber\\
  &\qquad+\varsigma(\cos(Z_{t}-Z_{t_{0}})\cos Z_{t_{0}}-\sin(Z_{t}-Z_{t_{0}})\sin Z_{t_{0}}+1).
\end{align}
The first integral on the right-hand side of (\ref{2.2.17}) clearly indicates that $V$ cannot be a Markov process or a semimartingale in general. In fact, it is so if and only if $h$ is chosen such that $h(t,s)-e^{-\kappa(t-t_{0})}h(t_{0},s)\equiv0$, a clear contradiction with the inclusion of short-term dependence; for instance, with the aforementioned three types of kernels this integral does not vanish. With the loss of the Markov property, it is oftentimes more comfortable to work directly with (\ref{2.2.16}). The conditional mean of the instantaneous variance can then be directly written down. For any fixed $t>t_{0}$ and $u>0$, we have
\begin{align}\label{2.2.18}
  \E[V_{t}|\mathscr{F}_{t_{0}}]&=V^{\circ}_{0}e^{-\kappa t}+\bar{V}(1-e^{-\kappa t})+\int^{t_{0}}_{0}h(t,s)\dd X_{s}+\xi_{1}\int^{t}_{t_{0}}h(t,s)\dd s \nonumber\\
  &\quad+\varsigma(\E[\cos Z_{t-t_{0}}]\cos Z_{t_{0}}-\E[\sin Z_{t-t_{0}}]\sin Z_{t_{0}}+1),
\end{align}
using that both $X$ and $Z$ have independent stationary increments.

\subsection{Average forward volatility}\label{sec:2.3}

After constructing the instantaneous variance model with roughness and jumps, we proceed to giving an explicit structure for the forward variance curve, i.e.,
\begin{equation}\label{2.3.1}
  \tilde{V}_{t}(u):=\E[V_{t+u}|\mathscr{F}_{t}],\quad u>0,\;t\geq0,
\end{equation}
which in light of (\ref{2.2.18}) admits the following stochastic representation,
\begin{align}\label{2.3.2}
  \tilde{V}_{t}(u)=\tilde{V}^{\circ}_{t}(u)+U_{t}(u),\quad u>0,\;t\geq0.
\end{align}
On the right-hand side of (\ref{2.3.2}),
\begin{equation}\label{2.3.3}
  \tilde{V}^{\circ}_{t}(u)=V^{\circ}_{0}e^{-\kappa(t+u)}+\bar{V}(1-e^{-\kappa(t+u)})+\int^{t}_{0}h(t+u,s)\dd X_{s}+\xi_{1}\int^{t+u}_{t}h(t+u,s)\dd s
\end{equation}
is the forward variance rising from the generalized fractional Ornstein-Uhlenbeck process $V^{\circ}$, whose fractional integral is associated with the shifted kernel $h(t+u,s)$, and
\begin{equation}\label{2.3.4}
  U_{t}(u)=\varsigma(\E[\cos Z_{u}]\cos Z_{t}-\E[\sin Z_{u}]\sin Z_{t}+1)
\end{equation}
stems from the composite-sinusoidal L\'{e}vy process that contains small two-sided volatility jumps.

Apart from the contemporaneous instantaneous variance, frictions in the forward variance curve also result from a new fractional L\'{e}vy process $X^{(H_{u})}$ containing additional information over the entire variance history. In consequence, with a general kernel $h$ the Markov property of $\tilde{V}(u)$ is completely lost. If one prefers to view $t+u>t$ as being time-independent, then (\ref{2.3.2}) really gives a martingale dynamics for the forward variance curve over $[0,t+u]$, which is similar to the martingale framework developed in [Horvath, 2020, \text{Sect.} 3] \cite{HJT}. However, we deliberately refrain from operating on such a framework as it is primarily beneficial from a simulation-based viewpoint. It is also worth emphasizing that, since $\lim_{s\nearrow t}H_{u}(t,s)=h(t+u,t)=O(1)$, $\forall t,u>0$, the modified process $X^{(H_{u})}$ deprives the sample paths of $\tilde{V}(u)$ of any degree of roughness, restoring the semimartingale property of $\tilde{V}(u)$.

Under continuous monitoring over a fixed window $\D>0$, the average forward volatility process is identified as the square root of the $\D$-running average of the forward variance process,
\begin{equation*}
  I_{t}(\D):=\sqrt{\frac{1}{\D}\int^{\D}_{0}\tilde{V}_{t}(u)\dd u},\quad t\geq0.
\end{equation*}
In the case of the VIX index, for instance, $\D=6/73$ year (equivalent to 30 days). By using (\ref{2.3.2}), some direct calculations lead to an integral representation of the corresponding average forward variance as well. For any $\D>0$ we have using the Fubini-Tonelli theorem that
\begin{align}\label{2.3.5}
  I^{2}_{t}(\D)\equiv\frac{1}{\D}\int^{\D}_{0}\tilde{V}_{t}(u)\dd u&=\frac{V^{\circ}_{0}(e^{-\kappa t}-e^{-\kappa(t+\D)})}{\kappa\D}+\bar{V}\bigg(1-\frac{e^{-\kappa t}-e^{-\kappa(t+\D)}}{\kappa\D}\bigg)+X^{(H_{\D})}_{t}+\xi_{1}\U_{\D}(t) \nonumber\\
  &\qquad+\frac{\varsigma}{\D}\bigg(\int^{\D}_{0}\E[\cos Z_{u}]\dd u\cos Z_{t}-\int^{\D}_{0}\E[\sin Z_{u}]\dd u\sin Z_{t}+\D\bigg),
\end{align}
where
\begin{equation}\label{2.3.6}
  H_{\D}(t,s):=\frac{1}{\D}\int^{\D}_{0}h(t+u,s)\dd u
\end{equation}
is a $\D$-forward integrated kernel. By construction $H_{\D}$ is continuously differentiable and both $H_{\D}$ and $\U_{\D}$ preserve stationarity from $h$.

After some tedious calculations it can be deduced that, when $h$ is the type-I kernel (\ref{2.2.6}), then
\begin{equation}\label{2.3.7}
  H_{\D}(t,s)\equiv H_{\D}(t-s)=\frac{\Gf(d,\kappa(t-s))-\Gf(d,\kappa(t-s+\D))}{\kappa^{d}\D\Gf(d)}
\end{equation}
and
\begin{equation}\label{2.3.8}
  \U_{\D}(t)\equiv\U_{\D}=\frac{(\kappa\D-d)\Gf(d)-\kappa\D\Gf(d,\kappa\D)+\Gf(d+1,\kappa\D)}{\kappa^{d+1}\D\Gf(d)}.
\end{equation}
Similarly, for the type-II kernel (\ref{2.2.9}) with $d\in(1/2,1]$,
\begin{equation}\label{2.3.9}
  H_{\D}(t,s)\equiv H_{\D}(t-s)=\frac{e^{-\kappa(t-s+\D)}(\Gf(d)-e^{\kappa\D}(\Gf(d)-\Gf(d,-\kappa(t-s)))-\Gf(d,-\kappa(t-s+\D)))}{(-\kappa)^{d}\D\Gf(d)}
\end{equation}
and
\begin{equation}\label{2.3.10}
  \U_{\D}(t)\equiv\U_{\D}=\frac{\D^{d}-e^{-\kappa\D}(-\kappa)^{-d}d(\Gf(d)-\Gf(d,-\kappa\D))}{\kappa\D\Gf(d+1)}.
\end{equation}
For the type-III kernel in its specialized form (\ref{2.2.15}) with $d\in(1/2,1)$, we have
\begin{align}\label{2.3.11}
  H_{\D}(t,s)&\equiv H_{\D}(t-s) \nonumber\\
  &=
  \begin{cases}
    \displaystyle \frac{(t-s+\D)^{d}-(t-s)^{d}}{\D\Gf(d+1)}&\quad\displaystyle \text{if }t-s+\D<\frac{1-d}{\kappa},\\
    \displaystyle \frac{((1-d)/\kappa)^{d}-(t-s)^{d}}{\D\Gf(d+1)}+\frac{e^{-\kappa(t-s+\D)+1-d}-1}{\kappa^{d}\D(1-d)^{2-d}\Gf(d-1)}&\quad\displaystyle \text{if }t-s<\frac{1-d}{\kappa}\leq t-s+\D,\\
    \displaystyle -\frac{e^{-\kappa(t-s+\D)+1-d}(e^{\kappa\D}-1)}{\kappa^{d}\D(1-d)^{2-d}\Gf(d-1)}&\quad\displaystyle \text{if }t-s\geq\frac{1-d}{\kappa}
  \end{cases}
\end{align}
and
\begin{equation}\label{2.3.12}
  \U_{\D}(t)\equiv\U_{\D}=
  \begin{cases}
    \displaystyle \frac{\D^{d}}{\Gf(d+2)}&\quad\displaystyle \text{if }\D<\frac{1-d}{\kappa},\\
    \displaystyle \frac{1}{\kappa^{d+1}\D(1-d)^{1-d}}\bigg(\frac{e^{-\kappa\D+1-d}}{\Gf(d)}+\frac{\kappa\D(d+1)+d(d-3)}{\Gf(d+2)}\bigg)&\quad\displaystyle \text{if }\D\geq\frac{1-d}{\kappa}.
  \end{cases}
\end{equation}
Note that all three types of $\D$-forward integrated kernels are stationary with constant averaging functions. From a computational viewpoint, a benefit of the piecewise nature of the type-III kernel is that one does not need to deal with incomplete gamma functions,\footnote{With the type-III kernel one can also discover a closed-form formula for the conditional characteristic function of the fractional process $V^{\circ}$ and hence the partial forward variance $\tilde{V}^{\circ}(u)$, which facilitates derivatives pricing on the instantaneous or forward variances. Details are put into Appendix B as the interest of the present paper is more for derivatives on the average forward variance.} which are even complex-valued under the type-II kernel.

The previous formulae for the forward variance curve $\tilde{V}(u)$ and its average $I^{2}(\D)$, \text{esp.} (\ref{2.3.3}), (\ref{2.3.4}) and (\ref{2.3.5}), can be conveniently restated for any fixed time interval $[t_{0},t]\subseteq[0,T]$, so that their conditional distributions may be analyzed. In particular, it is straightforward to deduce that, for $u>0$,
\begin{equation*}
  \tilde{V}^{\circ}_{t}(u)=\tilde{V}_{t_{0}}(t-t_{0}+u)-\xi_{1}\int^{t}_{t_{0}}h(t+u,s)\dd s+\int^{t}_{t_{0}}h(t+u,s)\dd X_{s}
\end{equation*}
and
\begin{align*}
  U_{t}(u)&=\varsigma(\E[\cos Z_{u}](\cos(Z_{t}-Z_{t_{0}})\cos Z_{t_{0}}-\sin(Z_{t}-Z_{t_{0}})\sin Z_{t_{0}})\\
  &\qquad-\E[\sin Z_{u}](\sin(Z_{t}-Z_{t_{0}})\cos Z_{t_{0}}+\cos(Z_{t}-Z_{t_{0}})\sin Z_{t_{0}})+1),
\end{align*}
which together yield
\begin{align}\label{2.3.13}
  I^{2}_{t}(\D)&=\frac{1}{\D}\int^{t-t_{0}+\D}_{t-t_{0}}\tilde{V}^{\circ}_{t_{0}}(u)\dd u-\xi_{1}\int^{t}_{t_{0}}H_{\D}(t,s)\dd s+\int^{t}_{t_{0}}H_{\D}(t,s)\dd X_{s} \nonumber\\
  &\qquad+\frac{\varsigma}{\D}\bigg(\int^{\D}_{0}\E[\cos Z_{u}]\dd u(\cos(Z_{t}-Z_{t_{0}})\cos Z_{t_{0}}-\sin(Z_{t}-Z_{t_{0}})\sin Z_{t_{0}}) \nonumber\\
  &\qquad-\int^{\D}_{0}\E[\sin Z_{u}]\dd u(\sin(Z_{t}-Z_{t_{0}})\cos Z_{t_{0}}+\cos(Z_{t}-Z_{t_{0}})\sin Z_{t_{0}})+\D\bigg).
\end{align}
From the last representation it is also noted that an attendant effect of $U_{t}(u)$, apart from its original purpose of capturing small-scale volatility jumps, is that the resultant average forward variance contains risks that cannot be fully spanned by the forward variance curve, which will only depend on $Z$, particularly the quadrant of $e^{\ii Z}$.\footnote{We stress that such volatility risks are not an oddity and the same phenomena also prevail when one considers the classical framework of log-volatility without resorting to any geometric average approximation. In the present framework, the tolerance of such risks is directly tied to the auxiliary parameter $\varsigma>0$.}

At this point, we have all the necessary tools to give a comfortable integral formula for the conditional characteristic function of the average forward variance, as the following proposition expounds.

\begin{proposition}\label{prop:2}
In the setting of (\ref{2.3.5}), given any $0\leq t_{0}<t\leq T$ and $\D>0$, it holds that
\begin{align}\label{2.3.14}
  \phi_{I^{2}_{t}(\D)|t_{0}}(l):=\E\big[e^{\ii lI^{2}_{t}(\D)}\big|\mathscr{F}_{t_{0}}\big]&=\frac{1}{\pi}\exp\bigg(\ii lJ(t,t_{0},\D)+\int^{t}_{t_{0}}\log\phi_{X_{1}}(lH_{\D}(t,s))\dd s\bigg) \nonumber\\
  &\qquad\times\int_{\mathds{R}}\psi(l,x;t_{0},\D)\int^{\infty}_{0}\Re\big[e^{-\ii\ell x}\phi^{t-t_{0}}_{Z_{1}}(\ell)\big]\dd\ell\dd x,\quad l\in\mathds{R},
\end{align}
where $\phi_{Z_{1}}$ is recalled to be the characteristic function of the random variable $Z_{1}$,
\begin{equation*}
  J(t,t_{0},\D):=\frac{1}{\D}\int^{t-t_{0}+\D}_{t-t_{0}}\tilde{V}^{\circ}_{t_{0}}(u)\dd u-\xi_{1}\int^{t}_{t_{0}}H_{\D}(t,s)\dd s>0
\end{equation*}
is a kernel-modulated forward variance quantity, and
\begin{align*}
  \psi(l,x;t_{0},\D)&:=\exp\bigg(\frac{\ii l\varsigma}{\D}\bigg(\Re\bigg[\frac{\phi^{\D}_{Z_{1}}(1)-1}{\log\phi_{Z_{1}}(1)}\bigg](\cos x\cos Z_{t_{0}}-\sin x\sin Z_{t_{0}})\\
  &\qquad-\Im\bigg[\frac{\phi^{\D}_{Z_{1}}(1)-1}{\log\phi_{Z_{1}}(1)}\bigg](\sin x\cos Z_{t_{0}}+\cos x\sin Z_{t_{0}})+\D\bigg)\bigg).
\end{align*}
\end{proposition}

The application of Proposition \ref{prop:2} can be facilitated if the density function of the L\'{e}vy process $Z$, \text{esp.} the Fourier inverse $(1/\pi)\int^{\infty}_{0}e^{-\ii\ell x}\phi^{t-t_{0}}_{Z_{1}}(\ell)\dd\ell$, can be written explicitly, allowing the target characteristic function to involve up to two parallel (multiplied) numerical integrals. Unfortunately, this would be an impractical requirement in consideration of the empirical finding in [Todorov and Tauchen, 2011, \text{Sect.} 6] \cite{TT}, which suggests that the L\'{e}vy measure $\nu_{Z}$ of $Z$ typically has a Blumenthal-Getoor index around 1.78, and there is no commonly known L\'{e}vy process with such a feature and yet has a closed-form density function. Despite this, benefiting from the composite-sinusoidal structure, $Z$ need not even have a finite variance and we can pick $Z$ from the family of symmetric $\alpha$-stable processes with the simple characteristic function $\phi_{Z_{1}}(l)=e^{-|l|^{\alpha}}$, for $\alpha\approx1.78$. The associated L\'{e}vy measure in (\ref{2.2.3}) is then $\nu_{Z}(\dd z)=-\sec(\pi\alpha/2)/(2\Gf(-\alpha)|z|^{\alpha+1})\dd z$ for $z\in\mathds{R}\setminus\{0\}$. Also, since the scale of the composite-sinusoidal process is already incorporated into the parameter $\varsigma\in(0,1)$, it is redundant to introduce additional parameters to $Z$, and due to symmetry of the assumed stable distribution the effect of the composite-sinusoidal process is solely to generate a suitable activity level of the average forward variance while other large asymmetric movements are taken account of by the partial forward variance $\tilde{V}^{\circ}(u)$ resulted from the generalized fractional Ornstein-Uhlenbeck process $V^{\circ}$.

Leastwise, with Proposition \ref{prop:2} one can deduce pricing-hedging formulae for derivatives contracts written on the average forward variance, e.g., the squared VIX index, which will be explained in detail in Section \ref{sec:4}. Most importantly, although it is not possible to derive a similar formula for the characteristic function of the average forward volatility, we will demonstrate how this difficulty may be overcome for volatility derivatives by way of power-type extensions. Moreover, by forcing $\D\searrow0$ (\ref{2.3.14}) is nothing but the conditional characteristic function for the instantaneous variance, i.e., $\E\big[e^{\ii lV_{t}}\big|\mathscr{F}_{t_{0}}\big]$, $l\in\mathds{R}$.

\section{Simulation techniques}\label{sec:3}

Despite general non-stationarity, using (\ref{2.2.1}) we can still simulate the sample paths of the instantaneous variance process. To do this we discretize the generic global time interval $[0,T]$ by means of the uniform partition
\begin{equation}\label{3.1}
  \mathbf{T}_{M}:=\bigg\{\frac{nT}{M}\bigg\}^{M}_{n=0},\quad M\in\mathds{N}_{++},\; M\gg1.
\end{equation}
Then, using the Gauss quadrature rule ([Golub and Welsch, 1969] \cite{GW}) and the L\'{e}vy properties of $X$, the Volterra-type stochastic integral in the definition of $V^{\circ}$ in (\ref{2.2.1}) can be approximated by a finite random sum,
\begin{equation}\label{3.2}
  \check{V}^{\circ}_{nT/M}=V_{0}e^{-\kappa nT/M}+\bar{V}\big(1-e^{-\kappa nT/M}\big)+\sum^{n-1}_{k=0}h\bigg(\frac{nT}{M},\frac{kT}{M}\bigg)\check{X}_{k},\quad n\geq1,
\end{equation}
where $\check{X}_{k}$'s are \text{i.i.d.} random variables with characteristic function $\E\big[e^{\ii l\check{X}_{k}}\big]=(\phi_{X_{1}}(l))^{T/M}$, for $l\in\mathds{R}$. If the kernel is stationary, $h(nT/M,kT/M)=h((n-k)T/M)$. Then, we have the following estimator of (\ref{2.2.1}) on $\mathbf{T}_{M}$,
\begin{equation*}
  \check{V}_{nT/M}=\check{V}^{\circ}_{nT/M}+\varsigma\Bigg(\cos\sum^{n}_{k=1}\check{Z}_{k}+1\Bigg),\quad n\geq1,
\end{equation*}
where likewise $\check{Z}_{k}$'s are \text{i.i.d.} random variables with characteristic function $\E\big[e^{\ii l\check{Z}_{k}}\big]=(\phi_{Z_{1}}(l))^{T/M}$, for $l\in\mathds{R}$. The next proposition describes the convergence rate of the discretized process $\check{V}$ towards $V$ over $(0,T]$ (with trivial equivalence at time 0).

\begin{proposition}\label{prop:3}
Under $\mathbf{T}_{M}$, for any fixed $t\in(0,T]$, there exists $n\in\mathds{N}\cap[1,M]$ such that the estimator $\check{V}_{nT/M}$ is conditionally asymptotically unbiased towards $V_{t}$ and
\begin{equation*}
  \E\big[\big(\check{V}_{nT/M}-V_{t}\big)^{2}\big]=\Var[V_{t}]+O(M^{-1}),\quad\text{as }M\rightarrow\infty.
\end{equation*}
\end{proposition}

Note that the convergence rate is unaffected by the fraction index $d$ of $h$, which applies to the three types of kernels discussed before. Nonetheless, the above $L^{2}$-convergence fails in the limit as $d\searrow1/2$. In a similar fashion, we can use the representation (\ref{2.3.5}) to simulate the sample paths of the average forward variance $I^{2}(\D)$ for a given $\D>0$, by using the estimator
\begin{align*}
  \check{I}^{2}_{nT/M}(\D)&=\frac{V^{\circ}_{0}(e^{-\kappa nT/M}-e^{-\kappa(nT/M+\D)})}{\kappa\D}+\bar{V}\bigg(1-\frac{e^{-\kappa nT/M}-e^{-\kappa(nT/M+\D)}}{\kappa\D}\bigg) \\
  &\qquad+\sum^{n-1}_{k=0}H_{\D}\bigg(\frac{nT}{M},\frac{kT}{M}\bigg)\check{X}_{k}+\xi_{1}\U_{\D}(nT/M) \\
  &\qquad+\frac{\varsigma}{\D}\Bigg(\int^{\D}_{0}\E[\cos Z_{u}]\dd u\cos\sum^{n-1}_{k=1}\check{Z}_{k}-\int^{\D}_{0}\E[\sin Z_{u}]\dd u\sin\sum^{n-1}_{k=1}\check{Z}_{k}+\D\Bigg),\quad n\in\mathds{N}\cap[1,M],
\end{align*}
to which the $L^{2}$-convergence criterion in Proposition \ref{prop:3} naturally applies. Besides, there is no need to discretize the deterministic integrals containing the sinusoidal composites of $Z_{u}$ which can be directly computed according to Proposition \ref{prop:2}.

\section{Power-type derivatives}\label{sec:4}

In this section we present the main pricing-hedging formulae for European-style derivatives written on the adjusted average forward volatility. The setting of Section \ref{sec:2.2} is adopted throughout, and a fixed maturity date $T>0$ is assumed.

\subsection{Power swaps}\label{sec:4.1}

We start with swaps written on the average forward volatility. The payoff of a power swap to its investor is at $T$
\begin{equation}\label{4.1.1}
  S^{(p)}_{T}=I^{p}_{T}(\D),
\end{equation}
where $p\geq0$ is a predetermined power coefficient. Here we have disregarded the notional amount for simplicity as it is merely a positive scaling factor. Of course, the extremal case $p=0$ corresponds to a fixed cash payment of 1 dollar, while by choosing $p=1$ and $p=2$, one obtains the standard volatility swap and the standard variance swap, respectively.

Convention is that, at inception of trading, the price of the swap is set to achieve a zero fair value, and so the price of the swap at a given time point $t_{0}$ before maturity can be simply computed as the expected value of $I^{p}_{T}(\D)$ conditional on $\mathscr{F}_{t_{0}}$, i.e., as
\begin{equation*}
  S^{(p)}_{t_{0}}=\E\big[I^{p}_{T}(\D)\big|\mathscr{F}_{t_{0}}\big],
\end{equation*}
and is treatable as a potentially fractional moment of $I^{2}_{T}(\D)$.

\begin{proposition}\label{prop:4}
At time $t_{0}\in[0,T)$, the price of the power volatility swap with payoff (\ref{4.1.1}) satisfies the quasi-recurrence relation\footnote{For simplicity we use $\phi^{(\varpi)}_{t_{0},T}(l;\D)\equiv\pd^{\varpi}\phi_{I^{2}_{T}(\D)|t_{0}}(l)/\pd l^{\varpi}$ to denote the $\varpi$th derivative of $\phi_{I^{2}_{T}(\D)|t_{0}}(l)$ with respect to $l\in\mathds{R}$, for $\varpi\geq0$.}
\begin{align}\label{4.1.2}
  S^{(p)}_{t_{0}}&=(-\ii)^{p/2}\phi^{(p/2)}_{I^{2}_{T}(\D)|t_{0}}(0),\quad p\in2\mathds{N}, \nonumber\\
  \rightsquigarrow S^{(p)}_{t_{0}}&=\sec\frac{\pi(p/2-\lfloor p/2\rfloor)}{2}\frac{p/2-\lfloor p/2\rfloor}{\Gf(1-p/2+\lfloor p/2\rfloor)}\nonumber\\
  &\qquad\times\int^{\infty}_{0}\Re\Bigg[\frac{S^{(2\lfloor p/2\rfloor)}_{t_{0}}-(-\ii)^{\lfloor p/2\rfloor}\phi^{(\lfloor p/2\rfloor)}_{I^{2}_{T}(\D)|t_{0}}(l)}{l^{p/2-\lfloor p/2\rfloor+1}}\Bigg]\dd l,\quad p\notin2\mathds{N},
\end{align}
provided that $\E\big[I^{p}_{T}(\D)\big]<\infty$.
\end{proposition}

Due to the exponential structure (\ref{2.3.14}) the convergence of (\ref{4.1.2}) is directly linked to the smoothness of the characteristic function of $X_{1}$ and $Z_{1}$ (refer to (\ref{2.1.5}) and (\ref{2.2.3})). Comprehensibly, it is far from exorbitant to demand that $\phi_{I^{2}_{T}(\D)|t_{0}}(\cdot)\in\mathcal{C}^{\lfloor p/2\rfloor+1}(\mathds{R})$, which condition remains valid for a wide class of square-integrable L\'{e}vy processes $X$. For instance, for $X$ a tempered stable process and $Z$ a stable process, its characteristic function (\ref{2.1.5}) immediately renders $\phi_{I^{2}_{T}(\D)|t_{0}}(\cdot)\in\mathcal{C}^{\infty}(\mathds{R})$ so that Proposition \ref{prop:4} is automatically applicable for all values of $p\geq0$. Implementation of (\ref{4.1.2}) is also nowhere near computationally intense, regardless of specializations of the ($\D$-)forward integrated kernels, by means of the Gauss quadrature rule for numerical integration and finite-difference approximations for differentiation of integer orders; for example, given the required degree of smoothness, a central approximation reads for $\epsilon>0$ small
\begin{equation*}
  \phi^{(\lfloor p/2\rfloor)}_{I^{2}_{T}(\D)|t_{0}}(0)=\sum^{\lfloor p/2\rfloor}_{n=0}\binom{\lfloor p/2\rfloor}{n}(-1)^{n}\frac{\phi_{I^{2}_{T}(\D)|t_{0}}\bigg(\bigg(\frac{\lfloor p/2\rfloor}{2}-n\bigg)\epsilon\bigg)}{\epsilon^{\lfloor p/2\rfloor}}+O(\epsilon^{2}).
\end{equation*}

In particular, by taking $p=1$ in (\ref{4.1.2}) the pricing formula for the standard volatility swap reads
\begin{equation}\label{4.1.3}
  S^{(1)}_{t_{0}}=\frac{1}{\sqrt{2\pi}}\int^{\infty}_{0}\Re\bigg[\frac{1-\phi_{I^{2}_{T}(\D)|t_{0}}(l)}{\sqrt{l^{3}}}\bigg]\dd l,
\end{equation}
whilst that for the corresponding variance swap is none but the $\mathscr{F}_{t_{0}}$-conditional mean of $I^{2}_{T}(\D)$ and we recall (\ref{2.3.1}) through (\ref{2.3.4}).

Amidst a non-Markovian setting, it is unrealistic to construct a perfect hedge for these power volatility swaps based on the forward variance curve only. In light of the structure of the characteristic function (\ref{2.3.14}), the dominant parts of $I^{2}(\D)$ not containing the active small-scale jumps of $\cos Z$ can leastways be hedged perfectly with the partial forward variance curve $\tilde{V}^{\circ}(u)$, \text{esp.} the kernel-modulated forward variance $J(T,t_{0},\D)$. In other words, we wish to find a partial hedging strategy designated for those dominant parts over the time period $[t_{0},T)$ which will require the entire forward variance curve $\{\tilde{V}_{t_{0}}(u):u\in(T-t_{0},T-t_{0}+\D]\}$, whereas other non-hedged volatility risks are precisely those that cannot be spanned and are all buried in the bounded sinusoidal composites involving the randomness of $Z$ exclusively. To that end we first define the time-indexed differential operator
\begin{equation*}
  \triangle_{t}:=\frac{\pd}{\pd J(t,t_{0},\D)},\quad t\in(t_{0},T],
\end{equation*}
for a fixed $t_{0}\in[0,T)$.

\begin{corollary}\label{cor:1}
In the setting of Proposition \ref{prop:4} we have\footnote{Although the notation $\phi^{(-1)}_{t_{0},T}(\cdot;\D)$ can be well understood as an antiderivative, it does not matter here due to multiplication by 0.}
\begin{align}\label{4.1.4}
  \triangle_{T}\big(S^{(p)}_{t_{0}}\big)&=\frac{p(-\ii)^{p/2-1}\phi^{(p/2-1)}_{I^{2}_{T}(\D)|t_{0}}(0)}{2}=\frac{pS^{(p-2)}_{t_{0}}}{2},\quad p\in2\mathds{N}, \nonumber\\
  \rightsquigarrow\triangle_{T}\big(S^{(p)}_{t_{0}}\big)&=\sec\frac{\pi(p/2-\lfloor p/2\rfloor)}{2}\frac{p/2-\lfloor p/2\rfloor}{\Gf(1-p/2+\lfloor p/2\rfloor)}\int^{\infty}_{0}\Re\bigg[\frac{1}{l^{p/2-\lfloor p/2\rfloor+1}}\bigg(\triangle_{T}\big(S^{(2\lfloor p/2\rfloor)}_{t_{0}}\big)\nonumber\\
  &\qquad-(-\ii)^{\lfloor p/2\rfloor-1}\bigg(l\phi^{(\lfloor p/2\rfloor)}_{I^{2}_{T}(\D)|t_{0}}(l)+\bigg\lfloor\frac{p}{2}\bigg\rfloor\phi^{(\lfloor p/2\rfloor-1)}_{I^{2}_{T}(\D)|t_{0}}(l)\bigg)\bigg)\bigg]\dd l,\quad p\notin2\mathds{N}.
\end{align}
\end{corollary}

The first equation in (\ref{4.1.4}) signifies that if $p/2$ is an integer, then the size of the (partial) hedge is equal to the spot price of another power variance swap, with decremented power $p/2-1$. Again, after taking $p=1$ in the second equation, we find the hedge for the standard volatility swap as
\begin{equation*}
  \triangle_{T}\big(S^{(1)}_{t_{0}}\big)=\frac{1}{\sqrt{2\pi}}\int^{\infty}_{0}\Re\bigg[\frac{\phi_{I^{2}_{T}(\D)|t_{0}}(l)}{\sqrt{l}}\bigg]\dd l.
\end{equation*}
Obviously, for the corresponding variance swap, the hedge is exactly $J(T,t_{0},\D)$, with $\triangle_{T}\big(S^{(2)}_{t_{0}}\big)=1$.

Moreover, we stress that the convergence of (\ref{4.1.4}) does not rely on integrability of the characteristic function, namely $\phi_{t_{0},T}(\cdot;\D)\in L^{1}(\mathds{R})$. In fact, although this condition is considerably benign and realistic, it is not necessary for the stated results to hold. The only assumption we have made in this regard is that $X_{1}$ and $Z_{1}$ are continuous random variables, connected with the non-atomic L\'{e}vy measures $\nu_{X}$ and $\nu_{Z}$.

\subsection{Asymmetric power options}\label{sec:4.2}

As mentioned since the introduction, our study for asymmetric power options is motivated by the average forward volatility being the square root of the average forward variance. In other words, an option written on the average forward volatility can be effectively treated as a power-type option on the average forward variance, with power exactly equal to $1/2$. For convenience and generality we still conduct our analysis subject to a positive power coefficient.

Let us consider a European-style put option contract on the average forward volatility $I_{T}(\D)$, having the terminal payoff
\begin{equation}\label{4.2.1}
  P^{(p_{1},p_{2},\rm(a))}_{T}=\big(K^{p_{2}}-I^{p_{1}}_{T}(\D)\big)^{+},
\end{equation}
where $K>0$ is the volatility strike and $p_{1},p_{2}\geq0$ are two predetermined power coefficients. We refer to this type as being asymmetric since the power imposed on the strike can differ from that on the average forward volatility. The payoff structure (\ref{4.2.1}) grants the option investor a leveraged view on the average forward volatility. Since $I_{T}(\D)$ takes values within the unit interval under normal conditions, $p_{1}>1$ actually reduces the option investor's risk exposure, other things equal, while $0\leq p_{1}<1$ expands it, which is the exact opposite of the case of equity options (see [Xia, 2019, \text{pp.} 119] \cite{X2}). Obviously, for any fixed $p_{2}\geq0$, $P^{(1,p_{2},\rm(a))}_{T}$ corresponds to the terminal payoff of the standard volatility put option, while $P^{(1,p_{2},\rm(a))}_{T}$ represents that of a standard put option on the average forward variance. In the case of a call option, we have
\begin{equation}\label{4.2.2}
  C^{(p_{1},p_{2},\rm(a))}_{T}=\big(I^{p_{1}}_{T}(\D)-K^{p_{2}}\big)^{+}.
\end{equation}

The following proposition is given for arbitrary-time pricing of the asymmetric power option.

\begin{proposition}\label{prop:5}
The price of the asymmetric power put option with terminal payoff (\ref{4.2.1}) at time $t_{0}\in[0,T)$ is given by
\begin{equation}\label{4.2.3}
  P^{(p_{1},p_{2},\rm(a))}_{t_{0}}=\frac{K^{p_{2}}}{2}-\frac{1}{\pi}\int^{\infty}_{0}\Re\bigg[\bigg(K^{p_{2}}e^{-\ii K^{2p_{2}/p_{1}}l}+\frac{\Gf(p_{1}/2+1)-\Gf(p_{1}/2+1,\ii K^{2p_{2}/p_{1}}l)}{(\ii l)^{p_{1}/2}}\bigg)\frac{\phi_{I^{2}_{T}(\D)|t_{0}}(l)}{\ii l}\bigg]\dd l.
\end{equation}
The price of the asymmetric power call option with terminal payoff (\ref{4.2.2}) at time $t_{0}\in[0,T)$ is given by
\begin{equation}\label{4.2.4}
  C^{(p_{1},p_{2},\rm(a))}_{t_{0}}=P^{(p_{1},p_{2},\rm(a))}_{t_{0}}-K^{p_{2}}+S^{(p_{1})}_{t_{0}},
\end{equation}
where $S^{(p_{1})}_{t_{0}}$ is the contemporaneous price of a power swap on $I_{T}(\D)$ specified in Proposition \ref{prop:4}.
\end{proposition}

By taking $p_{1}=p_{2}=1$ one has the pricing formulae for the standard volatility options. In particular,
\begin{equation}\label{4.2.5}
  P^{(1,1,\rm(a))}_{t_{0}}=\frac{K}{2}-\frac{1}{\pi}\int^{\infty}_{0}\Re\bigg[\bigg(Ke^{-\ii K^{2}l}+\frac{\sqrt{\pi}/2-\Gf(3/2,\ii K^{2}l)}{\sqrt{\ii l}}\bigg)\frac{\phi_{I^{2}_{T}(\D)|t_{0}}(l)}{\ii l}\bigg]\dd l
\end{equation}
and, recalling (\ref{4.1.3}),
\begin{equation}\label{4.2.6}
  C^{(1,1,\rm(a))}_{t_{0}}=\frac{1}{\pi}\int^{\infty}_{0}\Re\bigg[\sqrt{\frac{\pi}{2l^{3}}}-\bigg(Ke^{-\ii K^{2}l}+\frac{\ii\sqrt{\pi}/2-\Gf(3/2,\ii K^{2}l)}{\sqrt{\ii l}}\bigg)\frac{\phi_{I^{2}_{T}(\D)|t_{0}}(l)}{\ii l}\bigg]\dd l-\frac{K}{2}.
\end{equation}
On the other hand, for the standard put option on the average forward variance with $p_{1}=2$, there is a significant reduction,
\begin{equation*}
  P^{(2,1,\rm(a))}_{t_{0}}=\frac{K}{2}-\frac{1}{\pi}\int^{\infty}_{0}\Re\bigg[\frac{(e^{-\ii Kl}-1)\phi_{I^{2}_{T}(\D)|t_{0}}(l)}{l^{2}}\bigg]\dd l.
\end{equation*}
The formulae (\ref{4.2.5}) and (\ref{4.2.6}) for the standard volatility options can be implemented with substantial efficiency provided that the conditional characteristic function takes the form of (\ref{2.3.14}) and facilitate calibration of the model on standard option prices.

Hedging of the asymmetric power options resembles that of the corresponding power swap, which only makes use of the partial forward variance curve. For the following we adopt the differential operator $\triangle_{t}$ for $t\in(t_{0},T]$.

\begin{corollary}\label{cor:2}
In the setting of Proposition \ref{prop:5}, hedges can be constructed as
\begin{equation}\label{4.2.7}
  \triangle_{T}\big(P^{(p_{1},p_{2},\rm(a))}_{t_{0}}\big)=-\frac{1}{\pi}\int^{\infty}_{0}\Re\bigg[\bigg(K^{p_{2}}e^{-\ii K^{2p_{2}/p_{1}}l}+\frac{\Gf(p_{1}/2+1)-\Gf(p_{1}/2+1,\ii K^{2p_{2}/p_{1}}l)}{(\ii l)^{p_{1}/2}}\bigg)\phi_{I^{2}_{T}(\D)|t_{0}}(l)\bigg]\dd l
\end{equation}
and
\begin{equation}\label{4.2.9}
  \triangle_{T}\big(C^{(p_{1},p_{2},\rm(a))}_{t_{0}}\big)=\triangle_{T}\big(P^{(p_{1},p_{2},\rm(a))}_{t_{0}}\big) +\triangle_{T}\big(S^{(p_{1})}_{t_{0}}\big),
\end{equation}
where $\triangle_{T}\big(S^{(p_{1})}_{t_{0}}\big)$ is as specified in Corollary \ref{cor:1}.
\end{corollary}

Once again, with the choice $p_{1}=p_{2}=1$, the standard volatility options can be hedged in terms of
\begin{equation*}
  \triangle_{T}\big(P^{(1,1,\rm(a))}_{t_{0}}\big)=-\frac{1}{\pi}\int^{\infty}_{0}\Re\bigg[\bigg(Ke^{-\ii K^{2}l}+\frac{\sqrt{\pi}/2-\Gf(3/2,\ii K^{2}l)}{\sqrt{\ii l}}\bigg)\phi_{I^{2}_{T}(\D)|t_{0}}(l)\bigg]\dd l
\end{equation*}
and
\begin{equation*}
  \triangle_{T}\big(C^{(1,1,\rm(a))}_{t_{0}}\big)=\frac{1}{\pi}\int^{\infty}_{0}\Re\bigg[\bigg(\frac{\ii\sqrt{\pi}/2-\Gf(3/2,\ii K^{2}l)}{\sqrt{\ii l}}-Ke^{-\ii K^{2}l}\bigg)\phi_{I^{2}_{T}(\D)|t_{0}}(l)\bigg]\dd l.
\end{equation*}

We remark that the hedging strategies constructed in Corollary \ref{cor:1} and Corollary \ref{cor:2} only target the dominant movements in the average forward volatility $I(\D)$ that are governed by the fractional process $X^{(h)}$, to which the prices of volatility derivatives will be sensitive, whereas the bounded fluctuations controlling the activity level of the $I(\D)$ remain un-hedged. Nevertheless, these partial hedges remain valid with or without rough volatility and can be made perfect when the activity controller process $Z$ is absent.

\subsection{Symmetric power options}\label{sec:4.3}

As in the case of equity options, the volatility option investor's risk exposure can also be adjusted by directly forcing a mutual power effect on the standard option payoff (similar to [Raible, 2000, \text{Sect.} 3.4] \cite{R1} and [Xia, 2019, \text{pp.} 120] \cite{X2}). This way of generalization understandably does not build any useful connection between options on the average forward volatility and the corresponding forward variance and is hence considered less important from the viewpoint of this paper's motivation. Nonetheless, for the sake of completeness and our interest we still provide a comprehensive analysis of the pricing-hedging methods for such so-called ``symmetric power options.''

In this connection let a European-style put option contract on $I_{T}(\D)$ have the following terminal payoff,
\begin{equation}\label{4.3.1}
  P^{(p,\rm(s))}_{T}=\big((K-I_{T}(\D))^{+}\big)^{p},
\end{equation}
where $K>0$ and $p\geq0$. In this structure both the strike price and the average forward volatility undergo the same power impact, and with binomial expansion we can rewrite
\begin{equation}\label{4.3.2}
  P^{(p,\rm(s))}_{T}=\sum^{\infty}_{k=0}\binom{p}{k}(-1)^{k}K^{p-k}I^{k}_{T}(\D)\1_{\{I_{T}(\D)<K\}} =\sum^{\infty}_{k=0}\binom{p}{k}(-1)^{k}K^{p-k}S^{(k)}_{T}\1_{\{I_{T}(\D)<K\}},
\end{equation}
which shows that, conditional on $\{I_{T}(\D)<K\}$, the symmetric put power option can be looked upon as a weighted sum of power volatility swaps, each associated with an integer power coefficient in $\mathds{N}$, which at $k=0$ is merely a cash payment of $K^{p}$. Clearly, (\ref{4.3.2}) is a finite sum if and only if $p\in\mathds{N}$.

\begin{figure}[H]
  \centering
  \includegraphics[width=2.9in]{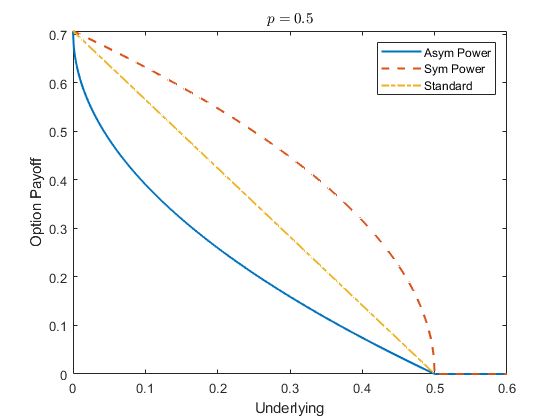}
  \includegraphics[width=2.9in]{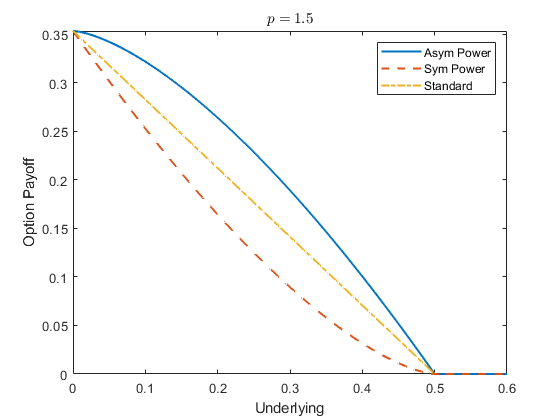}
  \caption{Comparison of leverage effects of power put options}
  \label{fig:2}
\end{figure}

Besides, we observe that the plots of the payoff functions $P^{(p,p,\rm(a))}_{T}$ and $P^{(p,\rm(s))}_{T}$ against $I_{T}(\D)$ are symmetric with respect to the line segment joining the points $(0,K^{p})$ and $(K,0)$ over the interval $[0,K]$. For $0\leq p<1$, the symmetric power option provides a convex transformation of the standard option payoff whereas its asymmetric power counterpart provides a concave one; for $p>1$ one has a reversed relation (see Figure \ref{fig:2}). Therefore, the two types of power put options can be utilized to complement each other in terms of severity of risk adjustment when either deeply in-the-money or closed to at-the-money. However, such effect holds exclusively for put options.

As we write the terminal payoff
\begin{align}\label{4.3.3}
  C^{(p,\rm(s))}_{T}&=\big((I_{T}(\D)-K)^{+}\big)^{p} \nonumber\\
  &=\begin{cases}
  \displaystyle \sum^{p}_{k=0}\binom{p}{k}(-K)^{k}S^{(p-k)}_{T}\1_{\{I_{T}(\D)>K\}},\quad p\in\mathds{N},\\
  \displaystyle \Bigg(\sum^{\lfloor p\rfloor}_{k=0}\binom{p}{k}(-K)^{k}S^{(p-k)}_{T}+\sum^{\infty}_{k=\lfloor p\rfloor+1}\binom{p}{k}(-K)^{k}I^{p-k}_{T}(\D)\Bigg)\1_{\{I_{T}(\D)>K\}},\quad p\notin\mathds{N},
  \end{cases}
\end{align}
a similar symmetric power call option can be decomposed into exactly $\lfloor p\rfloor$ weighted power volatility swaps incremented by an infinite sequence of power-type derivatives on the reciprocal average forward volatility $I^{-1}_{T}(\D)$ conditioned to stay below $1/K$, which vanishes if and only if $p$ is an integer. Also, note that in this case the payoff functions $C^{(p,p,\rm(a))}_{T}$ and $C^{(p,\rm(s))}_{T}$ are both strictly concave \text{resp.} convex in $I_{T}(\D)$ for $0\leq p<1$ \text{resp.} $p>1$ over $[K,\infty)$, with $\lim_{x\rightarrow\infty}(x^{p}-K^{p})^{+}/((x-K)^{+})^{p}=1$.

Based on the two decompositions (\ref{4.3.2}) and (\ref{4.3.3}), the next proposition gives the pricing formulae for these symmetric power options in terms of infinite series.

\begin{proposition}\label{prop:6}
The price of the symmetric power put option with terminal payoff (\ref{4.3.1}) at time $t_{0}\in[0,T)$ is given by
\begin{equation}\label{4.3.4}
  P^{(p,\rm (s))}_{t_{0}}=\frac{1}{\pi}\sum^{\infty}_{k=0}\binom{p}{k}(-1)^{k}K^{p-k}\int^{\infty}_{0}\Re\bigg[\frac{(\Gf(k/2+1)-\Gf(k/2+1,\ii K^{2}l))\phi_{I^{2}_{T}(\D)|t_{0}}(l)}{(\ii l)^{k/2+1}}\bigg]\dd l,
\end{equation}
while that of the similar symmetric power call option with (\ref{4.3.3}) is
\begin{align}\label{4.3.5}
  C^{p,\rm (s)}_{t_{0}}&=\sum^{\lfloor p\rfloor}_{k=0}\binom{p}{k}(-K)^{k}\bigg(S^{(p-k)}_{t_{0}} \nonumber\\
  &\qquad-\frac{1}{\pi}\int^{\infty}_{0}\Re\bigg[\frac{(\Gf((p-k)/2+1)-\Gf((p-k)/2+1,\ii K^{2}l))\phi_{I^{2}_{T}(\D)|t_{0}}(l)}{(\ii l)^{(p-k)/2+1}}\bigg]\dd l\bigg)+\Sigma^{(p)}_{t_{0}},
\end{align}
where $S^{(p-k)}_{t_{0}}$'s, for $0\leq k\leq p$, are the contemporaneous power swap prices as specified in Proposition \ref{prop:4} and
\begin{equation*}
  \Sigma^{(p)}_{t_{0}}=
  \begin{cases}
  \displaystyle 0, \quad&\text{if }p\in\mathds{N},\\
  \displaystyle \frac{1}{\pi}\sum^{\infty}_{k=\lfloor p\rfloor+1}\binom{p}{k}(-K)^{k}\int^{\infty}_{0}\Re\bigg[\frac{\Gf(1-(k-p)/2,\ii K^{2}l)\phi_{I^{2}_{T}(\D)|t_{0}}(l)}{(\ii l)^{1-(k-p)/2}}\bigg]\dd l,\quad&\text{if }p\notin\mathds{N}.
  \end{cases}
\end{equation*}
\end{proposition}

Hedges of these symmetric power options using $I^{2}_{t_{0}}(T-t_{0}+\D)$ also come in similar forms, which yield the next result.

\begin{corollary}\label{cor:3}
Assume the setting of Proposition \ref{prop:6}. Then we have
\begin{equation*}
  \triangle_{T}\big(P^{(p,\rm (s))}_{t_{0}}\big)=\frac{1}{\pi}\sum^{\infty}_{k=0}\binom{p}{k}(-1)^{k}K^{p-k}\int^{\infty}_{0}\Re\bigg[\frac{(\Gf(k/2+1)-\Gf(k/2+1,\ii K^{2}l))\phi_{I^{2}_{T}(\D)|t_{0}}(l)}{(\ii l)^{k/2}}\bigg]\dd l,
\end{equation*}
and
\begin{align*}
  \triangle_{T}\big(C^{p,\rm (s)}_{t_{0}}\big)&=\triangle_{T}\big(\Sigma^{(p)}_{t_{0}}\big)+\sum^{\lfloor p\rfloor}_{k=0}\binom{p}{k}(-K)^{k}\bigg(\triangle_{T}\big(S^{(p-k)}_{t_{0}}\big)\\
  &\qquad-\frac{1}{\pi}\int^{\infty}_{0}\Re\bigg[\frac{(\Gf((p-k)/2+1)-\Gf((p-k)/2+1,\ii K^{2}l))\phi_{I^{2}_{T}(\D)|t_{0}}(l)}{(\ii l)^{(p-k)/2}}\bigg]\dd l\bigg),
\end{align*}
where $\triangle_{T}\big(S^{(p-k)}_{t_{0}}\big)$'s, for $0\leq k\leq p$, are the contemporaneous power swap hedges as specified in Corollary \ref{prop:2} and
\begin{equation*}
  \triangle_{T}\big(\Sigma^{(p)}_{t_{0}}\big)=
  \begin{cases}
  \displaystyle 0, \quad&\text{if }p\in\mathds{N},\\
  \displaystyle \frac{1}{\pi}\sum^{\infty}_{k=\lfloor p\rfloor+1}\binom{p}{k}(-K)^{k}\int^{\infty}_{0}\Re\bigg[\frac{\Gf(1-(k-p)/2,\ii K^{2}l)\phi_{I^{2}_{T}(\D)|t_{0}}(l)}{(\ii l)^{(p-k)/2}}\bigg]\dd l,\quad&\text{if }p\notin\mathds{N}.
  \end{cases}
\end{equation*}
\end{corollary}

\section{An empirical study}\label{sec:5}

In this empirical study we illustrate the performance of our model framework established in Section \ref{sec:2} as well as the pricing-hedging formulae presented in Section \ref{sec:4}. Allowing for overall efficiency, we focus on the type-I and type-III kernels, specializing $h$ according to (\ref{2.2.6}) and (\ref{2.2.15}), respectively; as explained before, the type-II kernel (\ref{2.2.9}) derived from the Riemann-Liouville kernel can take negative values and does not possess an exponentially decaying right tail as demanded in (\ref{2.2.2}), and hence is likely to bring about numerical issues. The L\'{e}vy subordinator $X$ is taken to belong to the class of tempered stable subordinators, with characteristic function (\ref{2.1.5}), while the auxiliary L\'{e}vy process $Z$ is as mentioned in Section \ref{sec:2.3} a two-sided 1.78-stable process.\footnote{In adherence to how it is motivated, the stability index $\alpha$ is to be estimated from high-frequency data for the VIX index, along the lines of [Todorov and Tauchen, 2011, \text{Sect.} 4.2] \cite{TT}, rather than to be calibrated from option price data. In this context we simply stick to the value 1.78 in the wake of their empirical findings.}

\subsection{Data and preparation}\label{sec:5.1}

Since the VIX index is a popular gauge for the stock market volatility, it would be very interesting to evaluate the model performance under dissimilar market dynamics. For this reason, we have select two independent VIX option price data sets, in the years 2016 and 2020, which correspond to, respectively, normal times and the COVID-19 global pandemic -- it is known that in the latter period there has been significantly higher buying pressure into call options amid market fear, generating anomalous trading volumes. For both data sets, strike prices and option prices are quoted in the unit of US\$100 (data source: [CBOE Global Markets, Inc., 2020] \cite{CGM}) and we adopt $t_{0}=0$ throughout for simplicity.

In more detail, the first data set reflects ordinary market dynamics where the volatility smile is easily justified. It consists of 38 put option prices quoted on Jan 26th, 2016, under four different maturities $T=27,55,90,181$ days with the spot price $I_{0}(\D)=0.2667$ and the strike price $K$ ranging from 0.12 to 0.3. The second data set speaks to an abnormally volatile market dynamics with a conspicuous clustering of call option prices across different maturities. It contains 38 call option prices quoted as of May 11th, 2020, also corresponding to four maturities $T=72,100,163,191$ days with the spot price $I_{0}(\D)=0.3304$ and the strike price $K\in[0.2,0.9]$. It is clear that with these two data sets we can also simultaneously illustrate the pricing formulae for both call and put options.

By the definition of the VIX index we fix $\D=6/73$ (year). In order to apply Proposition \ref{prop:2} properly, one challenge that immediately comes to attention is the specification of the ($\mathscr{F}_{t_{0}}$-measurable) kernel-modulated forward variance $J(t,t_{0},\D)>0$, which arises because of the non-Markovian setting and contains all the information about the spot price of the VIX at time $t_{0}$. Understandably, it would be undesirable to treat the entire quantity $J(t,t_{0},\D)$ as an independent parameter to be calibrated, which can cause severe instability by disregarding the base level of the spot price. To overcome this difficulty, we employ an expansion argument to transform $J(t,t_{0},\D)$ into a linear combination of the square of the spot price (i.e., $I^{2}_{t_{0}}(\D)$) and a time-dependent remainder term with refined domains. Such an operation can significantly stabilize calibration by permitting the use of VIX index data. In particular, the relations (\ref{2.3.2}) and (\ref{2.3.4}) permit writing
\begin{equation}\label{5.1.1}
  J(t,t_{0},\D)=I^{2}_{t_0}(\D)-\xi_{1}\int^{t}_{t_{0}}H_{\D}(t,s)\dd s+r(t_{0},t),
\end{equation}
where the remainder, being bounded using the time-to-maturity and the scaling factor $\varsigma$,
\begin{equation}\label{5.1.2}
  |r(t_{0},t)|=\bigg|\frac{1}{\D}\int^{t}_{t_{0}}\big(\tilde{V}^{\circ}_{t_{0}}(v-t_{0}+\D)-\tilde{V}^{\circ}_{t_{0}}(v-t_{0})\big)(t-v)\dd v-\frac{1}{\D}\int^{\D}_{0}U_{t_{0}}(u)\dd u\bigg|\leq\frac{1}{\D}(t-t_{0})^2+3\varsigma,
\end{equation}
is to be calibrated independently in the domain $[-(t-t_{0})^2/\D-3\varsigma,(t-t_{0})^2/\D+3\varsigma]$. With (\ref{5.1.1}) and (\ref{5.1.2}), the corresponding formula (\ref{2.3.14}) for the characteristic function of the squared VIX is transformed into
\begin{align}\label{5.1.3}
  \phi_{I^{2}_{t}(\D)|t_{0}}(l)
  &=\frac{1}{\pi}\exp\bigg(\ii l\bigg(I^2_{t_0}(\D)-\xi_{1}\int^{t}_{t_{0}}H_{\D}(t,s)\dd s+r(t_{0},t)\bigg) +\int^{t}_{t_{0}}\log\phi_{X_{1}}(lH_{\D}(t,s))-\xi_1 H_{\Delta}(t,s)\dd s\bigg) \nonumber\\
  &\qquad\times\int_{\mathds{R}}\psi(l,x;t_{0},\D)\int^{\infty}_{0}\Re\big[e^{-\ii\ell x}\phi^{t-t_{0}}_{Z_{1}}(\ell)\big]\dd\ell\dd x,\quad l\in\mathds{R}.
\end{align}
The calibration of the reversion level $\bar{V}$ is automatically encoded into that of the remainder $r(t_{0},t)$. Again, we will specify the foregoing transformed formulae with $t_{0}=0$ and $t=T$ for implementation and write for simplicity $r(T)\equiv r(0,T)$.

\begin{figure}[H]
  \centering
  \includegraphics[width=2.7in]{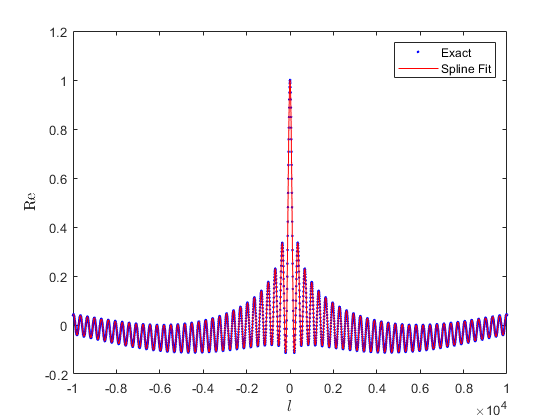}
  \includegraphics[width=2.7in]{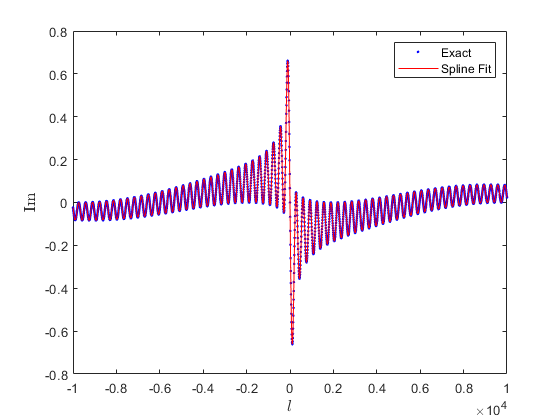}
  \includegraphics[width=2.7in]{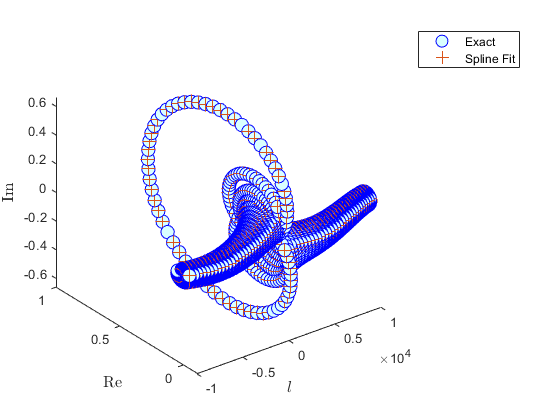}
  \caption{Smoothing spline approximation of $x$-integral}
  \label{fig:3}
\end{figure}

On a second look at the standard option pricing formulae (\ref{4.2.5}) and (\ref{4.2.6}), there are ultimately four numerical integrals to evaluate, having domains of integration $(0, \infty)\ni\ell$, $\mathds{R}\ni x$, $(0,T]\equiv(t_{0},t]\ni s$, and $(0,\infty)\ni l$, respectively. The first two form a repeated integral, and are then multipled by the third, whose product is nested with the fourth. This structure may seem intimidating at first glance; however, let us observe that: (i) the first integral $(1/\pi)\int^{\infty}_{0}\Re\big[e^{-\ii\ell x}\phi^{T}_{Z_{1}}(\ell)\big]\dd\ell$ is none but the probability density function of the 1.78-stable random variable $Z_{1}$, which are built-in in most software packages using efficient numerical Fourier inversion techniques; (ii) the scaling factor $\varsigma$ has limited contribution since its only purpose is to ensure that small volatility jumps have the desired activity index, 1.78; (iii) for a fixed small value (e.g., $0.01$) of $\varsigma$, the second integral with respect to $x$ can be approximated with arbitrary precision using a cubic smoothing spline thanks to $\psi$ being uniformly bounded by 1. With these observations in mind, we write $f_{Z_{T}}(x):= (1/\pi)\int^{\infty}_{0}\Re\big[e^{-\ii\ell x}\phi^{T}_{Z_{1}}(\ell)\big]\dd\ell$ and
proceed to fixing $\varsigma=0.01$ to carry out the following approximation of the second $x$-integral to boost computational efficiency,
\begin{equation}\label{5.1.4}
  \frac{1}{\pi}\int_{\mathds{R}}\psi(l,x;0,\D)\int^{\infty}_{0}\Re\big[e^{-\ii\ell x}\phi^{T}_{Z_{1}}(\ell)\big]\dd\ell\dd x= \int_{\mathds{R}}\psi(l,x;0,\D)f_{Z_{T}}(x)\dd x\approx\mathbf{f}_{\vec{\beta}}(l|\varsigma=0.01),
\end{equation}
where $\mathbf{f}_{\vec{\beta}}(l|\varsigma=0.01)$ is a cubic spine function defined with parameter $\vec{\beta} = (\beta_0, \beta_1, \beta_2, \beta_3)$ which is specified by solving the following minimization problem,
\begin{equation*}
  \vec{\beta} = \mathop{\text{argmin}}_{\vec{\beta}\in\mathds{R}^{4}} \sum^{n}_{k=1}\lambda \bigg(\int_{\mathds{R}}\psi(l_{k},x;t_{0},\D)f_{Z_{T}}(x)\dd x-\mathbf{f}_{\vec{\beta}}(l_{k})\bigg)^{2}+(1-\lambda) \int \big(\mathbf{f}_{\vec{\beta}}''(l)\big)^{2}\dd l,
\end{equation*}
where $n$ is the size of the partitioned $l$-domain and $\lambda\geq0$ is an auxiliary smoothing parameter. The role of the smoothing parameter is to balance between the fidelity to the true function and the roughness of the estimator by incorporating a penalty term linked to the curvature, and is set to be 0.0295 by default.\footnote{All implementation programs are written in MATLAB and are run using the scc2.bu.edu node on Boston University Shared Computing Cluster (SCC) with a detailed technical summary available at this \href{https://www.bu.edu/tech/support/research/computing-resources/tech-summary/}{link}.} The $l$-domain is partitioned by $n=3,000$ evenly distributed points over $[-10^{4}, 10^{4}]$, which is verified to be sufficient to produce a mean squared error estimate of $7.6\times10^{-11}$ and $8.2\times10^{-11}$ for the real and imaginary part, respectively. The effect of approximation for the real part, imaginary part and real-imaginary combined part are visualized in Figure \ref{fig:3}.

\subsection{Calibration exercise}\label{sec:5.2}

Calibration is carried out jointly taking into account all four different maturities for each complete data set. With the aid of (\ref{5.1.3}) and (\ref{5.1.4}), there are a total of six parameters to calibrate, $(a,b,c,d,\kappa,r(T))$, among which the last one with time dependence needs to be considered separately for different maturities and for this purpose we adopt the notation $T_{1}<T_{2}<T_{3}<T_{4}$. More specifically, the objective is to minimize the mean squared error (MSE) between the observed market prices of the VIX options and the corresponding model prices, so that the optimal parameter set is given by
\begin{equation}\label{5.2.1}
  \big(a,b,c,d,\kappa,\{r(T_{n})\}^{4}_{n=1}\big)=\underset{\substack{a>0,b>0,c\in(0,1),d\in(0.5,1),\kappa>0,\\r(T_{n})\in[-T^{2}_{n}/\D-3\varsigma, T^{2}_{n}/\D+3\varsigma],\\ n\in\{1,2,3,4\}|\alpha=1.78,\varsigma=0.01}}{\arg\min}\sum_{K,\{T_{n}\}^{4}_{n=1}}\big((\text{Market Prices})-(C,P)^{(1,1,\rm(a))}_{0}\big)^{2},
\end{equation}
where the sum runs over all available strike prices and maturities.

In order to solve (\ref{5.2.1}) efficiently, first we adopt the genetic algorithm, which conduces to locate a decent initial value $\vartheta_0$ of the parameters of interest. Then, we apply the pattern search algorithm taking $\vartheta_0$ as the initial value to continuously refine the parameter set. Both of these algorithms are parallelized so that the computation speed largely scales with the number of cores available.\footnote{In general, while the genetic algorithm search takes up to 8 hours on a 32-core processor, implementation of the pattern search with the initial value $\vartheta_0$ is much faster and the parameters can be refined within 2 hours. For this reason, for industry applications we recommend using the genetic algorithm to find the initial value $\vartheta_0$ once and for all, and subsequently implement the pattern search algorithm given $\vartheta_0$ as an initial point for daily updates.}

\begin{table}[H]\scriptsize
  \centering
  \caption{Calibration results (rounded to four decimal places)}
  \begin{tabular}{c|c|c|c|c|c|c|c|c|c|c}
  \hline
  Category & $a$ & $b$ & $c$ & $d$ & $\kappa$ & $r(T_1)$ & $r(T_2)$ & $r(T_3)$ & $r(T_4)$ & RMSE (\%) \\ \hline
  \begin{tabular}[c]{@{}c@{}}01/26/2016 puts\\ type-I kernel\end{tabular} & 0.1405 & 0.9269 & 0.5004 & 0.8994 & 3.0004 & 0.0044 & 0.0098  & 0.0165  & 0.0243  & 0.3497 \\ \hdashline
  \begin{tabular}[c]{@{}c@{}}01/26/2016 puts\\ type-III kernel\end{tabular} & 0.1378 & 1.6300 & 0.4351 & 0.7279 & 5.4844 & 0.0079 & 0.0118 & 0.0133 & 0.0108 & 0.2560 \\ \hline
  \begin{tabular}[c]{@{}c@{}}05/11/2020 calls \\ type-I kernel\end{tabular} & 0.3069 & 0.6716 & 0.6778 & 0.7226 & 6.0632 & 0.0695 & 0.0729  & 0.0655  & 0.0812  & 1.1636 \\ \hdashline
  \begin{tabular}[c]{@{}c@{}}05/11/2020 calls\\ type-III kernel\end{tabular} & 0.2979 & 1.8820 & 0.4732 & 0.5344 & 6.3233 & 0.0261 & 0.0359  & 0.0355  & 0.0459  & 1.0109 \\ \hline
  \end{tabular}
  \label{tab:1}
\end{table}

\begin{figure}[H]
  \centering
  \begin{minipage}[c]{0.49\linewidth}
  \centering
  \includegraphics[width=2.9in]{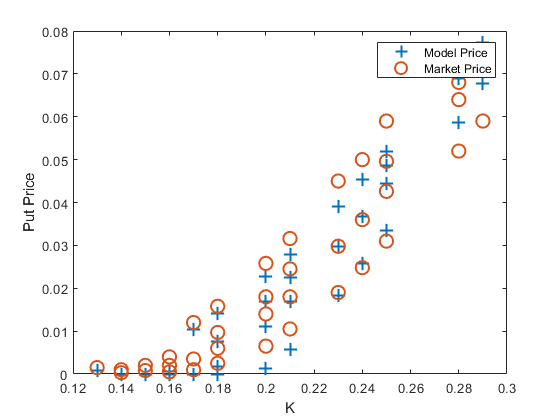}
  \caption*{\scriptsize 01/26/2016 puts (type-I kernel)}
  \includegraphics[width=2.9in]{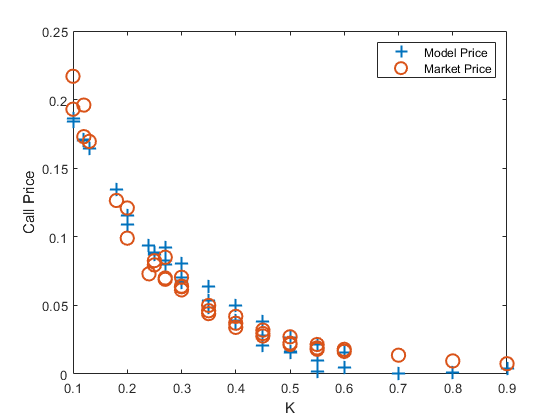}
  \caption*{\scriptsize 05/11/2020 calls (type-I kernel)}
  \end{minipage}
  \begin{minipage}[c]{0.49\linewidth}
  \centering
  \includegraphics[width=2.9in]{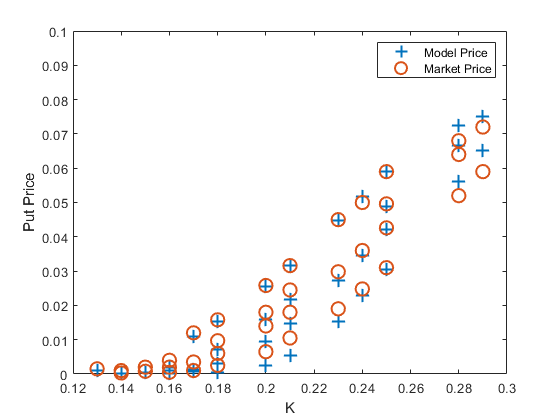}
  \caption*{\scriptsize 01/26/2016 puts (type-III kernel)}
  \includegraphics[width=2.9in]{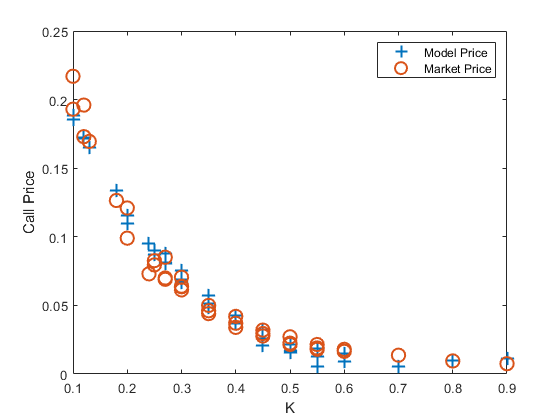}
  \caption*{\scriptsize 05/11/2020 calls (type-III kernel)}
  \end{minipage}
  \caption{VIX put and call option prices (market vs model)}
  \label{fig:4}
\end{figure}

Table \ref{tab:1} reports the calibrated parameter values for both data sets under both types of kernels, along with root mean squared errors (RMSE) at the level of the option prices. The model fits are further visualized in Figure \ref{fig:4} in four separate panels.

From Figure \ref{fig:4} it is immediately visible that during the COVID-19 pandemic market prices of call options are squeezed over different maturities, unlike the sparse price distribution of the put options in 2016. Despite this unusual discrepancy, the applied models have successfully captured the VIX option price styles for both periods. To be more precise, the calibration results seem promising for options with either short or long maturities, and for both in-the-money and deeply out-of-the-money options. From a comparative viewpoint, the model fits are much better for the first data set (pre-pandemic) which did not undergo abnormal trading volumes, and the same can be said about the model under the type-III kernel than the type-I. The latter observation may be partially explained by the piecewise construction of the type-III kernel, which precludes transcendental functions in the $\D$-forward integrated kernel ($H$) and thus facilitates numerical computations. As a result, the model under type-I kernel has performed relatively poorly with regard to those long-maturity options.

In interpreting the calibrated parameter values, from Table \ref{tab:1} it is clear that all the applied models are able to imply fast mean-reverting volatility, with $\kappa$ ranging from 3 to 6.4. The shape and scale parameters $a$ and $b$ of the base process $X$ are both close in value, speaking to the stability of calibration, while the family parameter $c$, with calibrated values around 0.5, suggests that a typical inverse Gaussian-driven (with $c=1/2$) Ornstein-Uhlenbeck process could be a suitable model for the instantaneous volatility process to capture volatility jumps. As for the fraction parameter $d$, since its value significantly differs from 1, a direct implication is that short-range dependence is indeed prevalent in volatility dynamics, even under unusual market environments during the pandemic.

\subsection{Power sensitivity analysis}\label{sec:5.3}

In this section we investigate the sensitivity of the pricing of power-type volatility derivatives for the VIX index with respect to varying power coefficients, by using the general formulae proposed in Section \ref{sec:4}. For succinctness we only look at one strike price $K=0.25$ and one maturity $T=90$ days for the first data set on put options and $K=0.35$ and $T =100$ days for the second on call options, adopting the parameter values calibrated under the type-III kernel in the third and fifth rows of Table \ref{tab:1}, respectively.

\begin{figure}[H]
  \centering
  \includegraphics[width=2.9in]{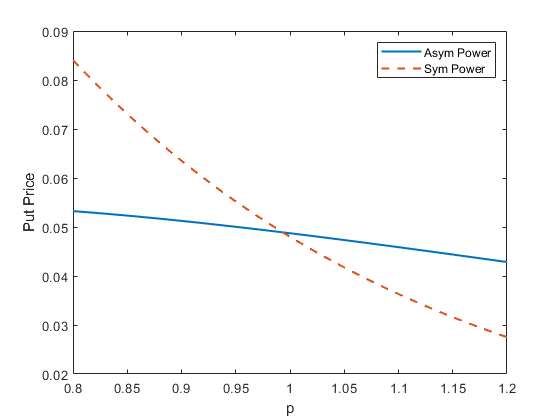}
  \includegraphics[width=2.9in]{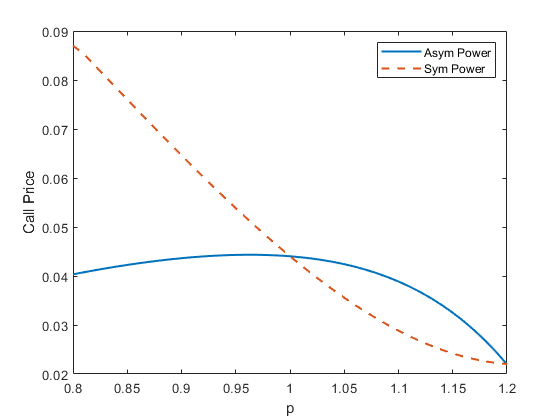}
  \caption{Power impact on VIX option prices}
  \label{fig:5}
\end{figure}

First, to ease comparison between asymmetric power and symmetric power types we assume for the power coefficients that $p_{1}=p_{2}=p\in[0.8,1.2]$ and plot the price changes ($C^{(p,p,{\rm (a)})}_{0}$, $P^{(p,p,{\rm (a)})}_{0}$, $C^{(p,{\rm (s)})}_{0}$ and $P^{(p,{\rm (s)})}_{0}$) of corresponding power-type derivatives in Figure \ref{fig:5}.

For the infinite series in Proposition \ref{prop:6} we use the approximation $\sum^{4}_{k=0}$ which universally leads to a global error less than $10^{-4}$. Plots for power swaps are excluded as they are already reflected in the call option pricing formulae thanks to the put-call parity. It is seen that, in terms of leverage exposure, the symmetric power options are able to provide much severer leverage effect for the VIX index compared to the asymmetric power options given the same power coefficients, despite that the latter are much easier to handle in general.

\begin{figure}[H]
  \centering
  \includegraphics[width=2.9in]{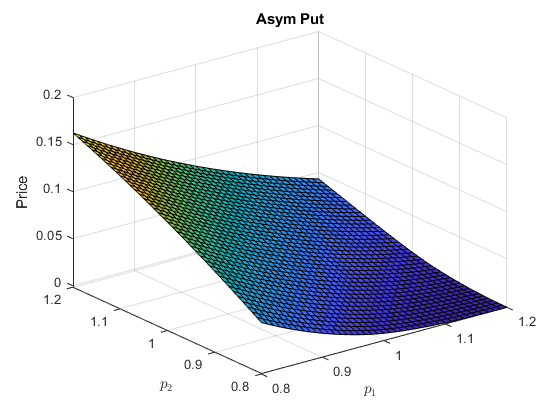}
  \includegraphics[width=2.9in]{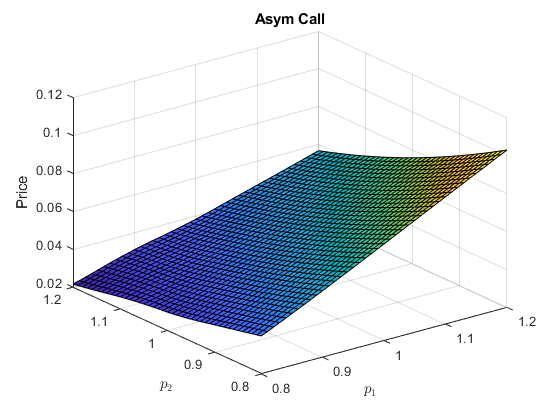}
  \caption{Asymmetric power surfaces for VIX option prices}
  \label{fig:6}
\end{figure}

Of course, for an asymmetric power option, by letting the two power coefficients vary independently we can generate a power surface for its price ($C^{(p_{1},p_{2},{\rm (a)})}_{0}$ and $P^{(p_{2},p_{2},{\rm (a)})}_{0}$), as shown in Figure \ref{fig:6}. Apart from showing the magnificent impact of powers on the VIX option price, these are also a reliable indicator that the general pricing-hedging formulae can be implemented fairly efficiently.

\section{Extension to rough volatility of volatility}\label{sec:6}

Needless to say, the discovery of [Da Fonseca and Zhang, 2019] \cite{DZ} provides yet another very interesting implication, that the volatility of the average forward volatility, such as the VVIX index, also exhibits short-term dependence. Although it is noticeably challenging to establish a comfortable framework coalescing both aspects of roughness, we will briefly discuss how the foregoing pricing problems may be tackled inheriting the structure of (\ref{2.2.1}). For that purpose we recall the setting of Section \ref{sec:2.3} and define the composite process
\begin{equation}\label{6.1}
  \tilde{I}_{t}(\D):=I_{\mathcal{T}_{t}}(\D),\quad t\geq0,
\end{equation}
with
\begin{equation}\label{6.2}
  \mathcal{T}_{t}:=\int^{t}_{0}Y^{(\eta)}_{s}\dd s,
\end{equation}
where $\mathcal{T}_{0}=0$ and $Y\equiv(Y_{t})$ is an $\mathbb{F}$-adapted square-integrable L\'{e}vy subordinator and $\eta$ a kernel which respectively resemble $X$ and $h$ up to different parameters. The construction (\ref{6.1}) emulates the initiative work of [Carr and Wu, 2004] \cite{CW} on stochastic time change, which has a fundamental root in the famous Dambis-Dubins-Schwartz theorem. Clearly, $Y^{(\eta)}=\int^{\cdot}_{0}\eta(\cdot,s)\dd Y_{s}$ is a mean-reverting process that introduces frictions into the volatility of the average forward volatility and, if $\eta$ is associated with a fraction parameter less than 1, say $d'\in(1/2,1]$, then Proposition \ref{prop:1} informs that $\sqrt{Y^{(\eta)}}$ also captures volatility-of-volatility jumps. By convention independence between $X$ and $Y$ is assumed.

Under (\ref{6.1}), we are able to at least write the unconditional characteristic function\footnote{Evaluating the conditional characteristic function on $\mathscr{F}_{t_{0}}$ for some $t_{0}\in[0,t)$ in the presence of time change can be cumbersome. Even if both $\eta$ and $h$ are exponential kernels posing no roughness, the time-changed process $\tilde{I}^{2}(\D)$ cannot be Markovian with respect to the filtration $\mathbb{F}$ jointly generated by $(X,Y,Z)$.} of the time-changed average forward variance in terms of nested integrals.

\begin{proposition}\label{prop:7}
Let $\phi_{Y_{1}}(l):=\E\big[e^{\ii lY_{1}}\big]$, $l\in\mathds{R}$, denote the characteristic function of $Y_{1}$. Then, for any $t>0$,
\begin{align}\label{6.3}
  \tilde{\phi}_{t}(l;\D)&:=\E\big[e^{\ii l\tilde{I}^{2}_{t}(\D)}\big] \nonumber\\
  &=\frac{1}{\pi}\int^{\infty}_{0}\phi_{0,y}(l;\D)\int^{\infty}_{0}\Re\bigg[\exp\bigg(-\ii\lambda y+\int^{t}_{0}\log\phi_{Y_{1}}\bigg(\lambda\int^{t}_{s}\eta(v,s)\dd v\bigg)\dd s\bigg)\bigg]\dd\lambda\dd y,\quad l\in\mathds{R},
\end{align}
where $\phi_{0,y}(l;\D)$ is as given in (\ref{2.3.14}).
\end{proposition}

Note that the innermost integral in (\ref{6.3}) can be expressed explicitly if $\eta$ is any of the three types of kernels specified before, while the other three integrals remain numerical in nature. In particular, the outer two integrals are generally non-interchangeable, i.e., Fubini's theorem is not applicable and they should be computed in sequence.

Regardless, we can put (\ref{6.3}) into the pricing formulae proposed in Section \ref{sec:4} to compute the prices of power volatility derivatives at time 0. However, since there are at least six numerical integrals (some being parallel) involved, coming up with a robust calibration scheme will be an arduous task. As a means of reducing calibration burden towards that end, one possibility is to conduct characteristic function-based estimation (see [Yu, 2004] \cite{Y}) based on volatility-of-volatility index data for the parameters of $Y$ beforehand.

\section{Concluding remarks}\label{sec:7}

The modeling of short-term dependence in instantaneous volatility is far from deep-rooted in Brownian sample paths and can be alternatively realized by way of a purely discontinuous L\'{e}vy process, which is able to capture volatility jumps as well. The latter approach is largely motivated from a balance between the two major empirical findings of [Todorov and Tauchen, 2011] \cite{TT} and [Gatheral et al, 2018] \cite{GJR}.

The pricing-hedging framework presented in this paper is tailored for volatility derivatives and built upon a generalized L\'{e}vy-driven Ornstein-Uhlenbeck process with a suitable square-integrable kernel to establish short-term dependence. An independent two-sided L\'{e}vy process in sinusoidal form is utilizable to help attain an empirically evidenced activity level of the average-forward variance. The elegance of the present framework lies in the unconditional positivity of the instantaneous variance, which eludes use of the logarithm, so that integration is largely facilitated leading eventually to a semi-closed characteristic function for the conditional forward variance. This signifies that there is no need of inexact transformations by geometric means (as in [Horvath et al, 2020] \cite{HJT} \text{e.g.}) whose levels of imprecision are, if possible at all, difficult to justify if there is no intention of running simulations. In other words, our approach is inherently analytical, on which basis various advanced numerical integration methods can be developed and directly applied. At the same time, we have discussed three types of kernels, the third of which, being completely new, is argued to be the most computation-friendly by involving only elementary functions and is hence recommended as an ideal substitute for the (commonly used) Riemann-Liouville kernel.

Since a volatility derivative is the same as a similar derivative written on the corresponding variance raised to the power $1/2$, it is natural to think of a wider class of power-type volatility derivatives also as a means of introducing leverage effect into the option payoffs. This is very much comparable to the original invention of power options written on equity or fixed-income instruments, in terms of functionalities. The general pricing-hedging formulae proposed in Section \ref{sec:4}, being entirely analytical in their own right, should be interpreted as model-independent requiring nothing more than a semi-closed characteristic function of the underlying quantity in order to operate. Besides, these new formulae can be thought of as power-exponential analogs of the well-known equity-option pricing formulae with exponential structures that initially appeared in [Bakshi and Madan, 2000] \cite{BM} -- more precisely, resultant exponential functions are mostly replaced by incomplete gamma functions.

Speaking of outcomes, our empirical study on VIX options has demonstrated that the proposed model framework and pricing formulae are generally highly stable and efficient for applications. Combined they are expected to fit considerably well for both short- and long-maturity, in- and deeply out-of-the-money options, and more so when calibrated under the type-III kernel, which is hence preferred over the first type. All the model parameters can be reliably calibrated under this framework, even including the fraction index $d$ and the tempered-stable family parameter $c$ that are tied to highly nonlinear relations, while the activity level of the VIX index is set to be $\alpha=1.78$ to allow for properly active small-scale fluctuations in the light of [Todorov and Tauchen, 2011]'s \cite{TT} discovery. The calibrated parameter values are in keeping with economic interpretations, confirming the prevalence of fast mean reversion ($\kappa$) and short-term dependence in volatility ($d$). The value of $c$ points out that large volatility jumps can be suitably dealt with by something close to an inverse Gaussian model, also signaling to some degree deviation from the exclusive use of a Brownian motion. In particular, the remarkable fits for the short-maturity options in the second data set, observed during the COVID-19 pandemic when trading volumes witnessed abnormal increases, would not likely have been achieved had (upward) volatility jumps been completely turned aside.

Overall, our methodology has provided new insights into how to think about the pricing problem of volatility derivatives allowing for both rough volatility and volatility jumps, by making full use of characteristic functions. With the foregoing remarks in mind, future research could be for instance devoted to understanding the advanced valuation of VIX derivatives under rough stochastic volatility of volatility (mentioning [Da Fonseca and Zhang, 2019] \cite{DZ} again), following the idea of Section \ref{sec:6} via a temporal composition process. The exploration of the significance of rough volatility in pricing derivatives linked to cryptocurrencies (mentioning [Takaishi, 2020] \cite{T1}) would also be an interesting subject.

\clearpage 

\section*{Appendix A - Proofs}

\subsection*{Proof of Proposition \ref{prop:1}}

For $V^{\circ}$ defined in (\ref{2.2.1}) we focus on the stochastic integral part, namely $X^{(h)}$, the rest being obviously continuous and of finite variation. Since $h$ is continuously differentiable and $h(t+u,t)=O\big(e^{-\kappa u}u^{(d-1)^{+}}\big)=o(u^{d-1})$ as $u\rightarrow\infty$ for any $t\geq0$, by the extreme value theorem there exist two positive constants $\mathfrak{b}_{h}\geq\mathfrak{a}_{h}>0$, which depend only on the parameters of $h$ including $\kappa$ and $d$, such that, for any $u>0$ and $t+u\in[0,T]$,
\begin{equation*}
  \E\big[\big(X^{(h)}_{t+u}-X^{(h)}_{t}\big)^{2}\big]\in\E\big[\big(X^{(g)}_{t+u}-X^{(g)}_{t}\big)^{2}\big]\times[\mathfrak{a}_{h},\mathfrak{b}_{h}], \tag{A.1.1}
\end{equation*}
where $g$ is the Riemann-Liouville kernel (\ref{2.1.3}) with the same fraction parameter $d$. This relation enables us to restrict our analysis to the Riemann-Liouville fractional L\'{e}vy subordinator $X^{(g)}$ without mean reversion. The expectation on the right-hand side of (A.1.1) is then by the L\'{e}vy-It\^{o} isometry
\begin{align*}
  \mathcal{E}^{(g)}_{t,u}:=\E\big[\big(X^{(g)}_{t+u}-X^{(g)}_{t}\big)^{2}\big] &=\frac{\xi_{2}}{\Gf^{2}(d)}\bigg(\int^{t}_{0}((t+u-s)^{d-1}-(t-s)^{d-1})^{2}\dd s+\int^{t+u}_{t}(t+u-s)^{2(d-1)}\dd s\bigg)\\
  &\qquad+\frac{\xi^{2}_{1}((t+u)^{d}-t^{d})^{2}}{\Gf^{2}(d+1)}.
\end{align*}
In particular, for $d>2$ there is the fundamental Riemann integral representation $X^{(g)}_{t}=\int^{t}_{0}\big(\int^{s}_{0}(s-v)^{d-2}/\Gf(d-1)\dd X_{v}\big)\dd s$ due to (\ref{2.1.4}) and so we only need to consider $d\leq2$.

Suppose $d\in(1,2]$. We observe that the uniformly continuous map
\begin{equation*}
  \mathds{R}_{++}\ni u\mapsto\frac{1}{\Gf^{2}(d)}\int^{t}_{0}((t+u-s)^{d-1}-(t-s)^{d-1})^{2}\dd s\in\mathds{R}_{++} \tag{A.1.2}
\end{equation*}
is increasing and convex for every $t>0$, hence the only need to compare its left tail behavior against power-law tails. It is not difficult to see that as $u\searrow0$,
\begin{equation*}
  \int^{t}_{0}((t+u-s)^{d-1}-(t-s)^{d-1})^{2}\dd s=O(u^{2}),
\end{equation*}
which implies that
\begin{equation*}
  \int^{t}_{0}((t+u-s)^{d-1}-(t-s)^{d-1})^{2}\dd s\leq\mathfrak{c}_{d}(T)u^{\min\{2d-1,2\}},\quad\forall t+u\in[0,T],\;u>0
\end{equation*}
for some constant $\mathfrak{c}_{d}(T)>0$ depending on $d$ and $T$. Also, $\int^{t+u}_{t}(t+u-s)^{2(d-1)}\dd s=u^{2d-1}/(2d-1)$ and combining things we claim that there exists another constant $\tilde{\mathfrak{c}}_{d}(T)>0$ such that
\begin{equation*}
  \mathcal{E}^{(g)}_{t,u}\leq\tilde{\mathfrak{c}}_{d}(T)u^{\min\{2d-1,2\}}. \tag{A.1.3}
\end{equation*}
Since the power of $u$ in (A.1.3) strictly exceeds 1, we apply the Kolmogorov-\v{C}entsov theorem (see, e.g., [Karatzas and Shreve, 1991, \text{Sect.} 2.2.B] \cite{KS}) to conclude that $X^{(g)}$, and hence $V$ due to (A.1.1), admits an \text{a.s.} continuous modification over $\mathds{R}_{+}$; in particular, the modification is guaranteed to be \text{a.s.} locally H\"{o}lder-continuous for every exponent in $(0,\min\{d-1,1/2\})$.

Furthermore, using the uniform time partition $\mathbf{T}_{M}$ (with $t_{0}=0$) in (\ref{3.1}) of the interval $[0,T]$, we define the $\mathbf{T}_{M}$-quadratic variation
\begin{equation*}
  Q_{M}([0,T]):=\sum^{M}_{n=1}\big(X^{(g)}_{nT/M}-X^{(g)}_{(n-1)T/M}\big)^{2}\geq0,\quad M\in\mathds{N}_{++},\; M\gg1.
\end{equation*}
If $d\in(1,2]$, by the convexity of (A.1.2) and the upper bound (A.1.3) we have
\begin{equation*}
  \E[Q_{M}([0,T])]\leq M\mathcal{E}^{(g)}_{(M-1)T/M,T/M}\leq\tilde{\mathfrak{c}}_{d}(T)\bigg(\frac{M}{T}\bigg)^{\max\{2(1-d),-1\}}\rightarrow0, \quad\text{as }M\rightarrow\infty,
\end{equation*}
which by nonnegativity and the relation (A.1.1) implies that $V$ has \text{a.s.} zero quadratic variation over $[0,T]$ and completes the proof of assertion (i).

Now suppose $d\in(1/2,1]$, let $X_{-}$ denote the c\`{a}gl\`{a}d modification of $X$ and set
\begin{equation*}
  E:=\{\omega\in\Omega:(X_{t}-X_{t-})(\omega)>0,\;\exists t\in[0,T]\}.
\end{equation*}
With $\nu(\mathds{R}_{++})=\infty$, it is a familiar result (see again [Lyasoff, 2017, \text{Sect.} 16] \cite{L2}) that $\PP E=1$. Then we observe that
\begin{equation*}
  X^{(h)}_{t}-X^{(h)}_{t-}=\int^{t-}_{0}(h(t,s)-h(t-,s))\dd X_{s}+\int^{t}_{t-}h(t,s)\dd X_{s},\quad t\geq0. \tag{A.1.4}
\end{equation*}
Since, for any $s\in[0,t)$, $\mathds{R}_{++}\ni t\mapsto h(t,s)\in\mathds{R}_{+}$ is a continuous map with $d\in(1/2,1]$, by the dominated convergence theorem the first integral in (A.1.4) is naught (in the sense of $L^{2}$-convergence) and
\begin{equation*}
  \int^{t}_{t-}h(t,s)\dd X_{s}=h(t,t-)(X_{t}-X_{t-}).
\end{equation*}
In consequence, there must exist some $t>0$ such that $X^{(h)}_{t}-X^{(h)}_{t-}\propto h(t,t-)$. To put it another way,
\begin{equation*}
  E\subseteq\big\{\omega\in\Omega:\big(X^{(h)}_{t}-X^{(h)}_{t-}\big)(\omega)\propto h(t,t-),\;\exists t\in[0,T]\big\}. \tag{A.1.5}
\end{equation*}
By (\ref{2.2.2}) further, if $d=1$ then $h$ is uniformly bounded so that $h(t,t-)>0$ for any $t>0$, and hence (A.1.5) proves the \text{a.s.} discontinuity of the sample paths of $X^{(h)}$. In this case, since $X^{(h)}$ has no Brownian part, its quadratic variation is given by the sum of its squared jumps,
\begin{equation*}
  \sum_{t\in[0,T]}\big(X^{(h)}_{t}-X^{(h)}_{t-}\big)^{2}=\sum_{t\in[0,T]}h^{2}(t,t-)(X_{t}-X_{t-})^{2}>0,\quad\PP\text{-a.s.},
\end{equation*}
where the inequality follows from the finiteness of $\sum_{t\in[0,T]}(X_{t}-X_{t-})^{2}>0$. On the other hand, if $d<1$, then $h(t,t-)=\infty$ for any $t>0$, which with (A.1.5) gives the \text{a.s.} discontinuity and unboundedness of the sample paths of $X^{(h)}$, and hence $V$, over $[0,T]$. An immediate implication is therefore that the sample paths of $V$ have infinitely large squared jumps over $[0,T]$ $\PP$-a.s., from which follows its ($\PP$-a.s.) infinite quadratic variation. Therefore assertions (ii) and (iii) are proved. $\square$

\subsection*{Proof of Proposition \ref{prop:2}}

According to the conditional representation (\ref{2.3.13}), we write for fixed $0\leq t_{0}<t\leq T$
\begin{equation*}
  I^{2}_{t}(\D)=J(t,t_{0},\D)+\int^{t}_{t_{0}}H_{\D}(t,s)\dd X_{s}+\log\psi(-\ii,Z_{t}-Z_{t_{0}};t_{0},\D), \tag{A.2}
\end{equation*}
where $J(t,t_{0},\D)$ is as defined in Proposition \ref{prop:2} and preliminarily
\begin{align*}
  \psi(l,Z_{t}-Z_{t_{0}};t_{0},\D)&:=\exp\bigg(\frac{\ii l\varsigma}{\D}\bigg(\int^{\D}_{0}\E[\cos Z_{u}]\dd u(\cos x\cos Z_{t_{0}}-\sin x\sin Z_{t_{0}})\\
  &\qquad-\int^{\D}_{0}\E[\cos Z_{u}]\dd u(\sin x\cos Z_{t_{0}}+\cos x\sin Z_{t_{0}})+\D\bigg)\bigg),\quad l\in\mathds{C}.
\end{align*}
Due to the infinite divisible distribution of $Z_{1}$, $\int^{\D}_{0}\E[\cos Z_{u}]\dd u=\Re\big[\int^{\D}_{0}\phi^{u}_{Z_{1}}(1)\dd u\big]=\Re\big[\big(\phi^{\D}_{Z_{1}}(1)-1\big)\big/\log\phi_{Z_{1}}(1)\big]$ and similarly $\int^{\D}_{0}\E[\sin Z_{u}]\dd u=\Im\big[\big(\phi^{\D}_{Z_{1}}(1)-1\big)\big/\log\phi_{Z_{1}}(1)\big]$. On the right-hand side of (A.2) $\int^{t}_{t_{0}}H_{\D}(t,s)\dd X_{s}$ is independent from $\mathscr{F}_{t_{0}}$. Because the process $\int^{\cdot}_{t_{0}}H_{\D}(t,s)\dd X_{s}$ is additive (or has independent increments) on $(t_{0},T]$ with $t$ treated as fixed and $Z$ is a L\'{e}vy process having an absolutely continuous distribution, using in proper order the independence lemma (for $X$ and $Z$), the infinite divisibility of the distribution of $X_{1}$, and the inverse Fourier transform, we have for $l\in\mathds{R}$
\begin{align*}
  \E\big[e^{\ii lI^{2}_{t}(\D)}\big|\mathscr{F}_{t_{0}}\big]&=e^{\ii lJ(t,t_{0},\D)}\E\big[e^{\ii l\int^{t}_{t_{0}}H_{\D}(t,s)\dd X_{s}}\big]\E\big[\psi(l,Z_{t}-Z_{t_{0}};t_{0},\D)|\mathscr{F}_{t_{0}}\big]\\
  &=e^{\ii lJ(t,t_{0},\D)}\prod^{t}_{t_{0}}\E\big[e^{\ii lH_{\D}(t,s)X_{1}}\big]^{\dd s}\int_{\mathds{R}}\psi(l,x;t_{0},\D)f_{Z_{t-t_{0}}}(x)\dd x\\
  &=e^{\ii lJ(t,t_{0},\D)}\bigg(\exp\int^{t}_{t_{0}}\log\E\big[e^{\ii lH_{\D}(t,s)X_{1}}\big]\dd s\bigg)\int_{\mathds{R}}\psi(l,x;t_{0},\D)\frac{1}{\pi}\int^{\infty}_{0}\Re\big[e^{-\ii\ell x}\phi^{t-t_{0}}_{Z_{1}}(\ell)\big]\dd\ell\dd x,
\end{align*}
where $\prod^{\cdot}_{\cdot}$ stands for the geometric integral operator (see, e.g., [Slav\'{i}k, 2007] \cite{S2}) and $f_{Z_{t-t_{0}}}(x)$, $x\in\mathds{R}$, denotes the density function of the random variable $Z_{t-t_{0}}$. After rearrangement the desired integral representation (\ref{2.3.14}) is therefore obtained. $\square$

\subsection*{Proof of Proposition \ref{prop:3}}

First notice that
\begin{equation*}
  \E\big[\check{V}_{nT/M}\big]=V_{0}e^{-\kappa nT/M}+\bar{V}\big(1-e^{-\kappa nT/M}\big)+\xi_{1}\sum^{n-1}_{k=0}h\bigg(\frac{nT}{M},\frac{kT}{M}\bigg)\frac{T}{M} +\varsigma\Bigg(\E\Bigg[\cos\sum^{n}_{k=1}\check{Z}_{k}\Bigg]+1\Bigg).
\end{equation*}
For a given $t\in(0,T]$, we choose $n\equiv n(t,M)=\lfloor Mt/T\rfloor$, so that $\lim_{M\rightarrow\infty}(n(t,M)T/M)=t$. Since the Riemann integral $\int^{t}_{0}h(t,s)\dd s$ is well-defined for $t\in[0,T)$, (\ref{3.2}) constitutes a conventional rectangular Riemann sum approximation and it is familiar that
\begin{equation*}
  \E\big[\check{V}^{\circ}_{n(t,M)T/M}-V^{\circ}_{t}\big]=O(M^{-1}),\quad\text{as }M\rightarrow\infty.
\end{equation*}
Also, $\E\big[\cos\sum^{n(t,M)}_{k=1}\check{Z}_{k}-\cos Z_{t}\big]=O(M^{-1})$ as the cosine is continuous differentiable. These show asymptotic unbiasedness. In addition, using the relation
\begin{equation*}
  \E\big[\big(\check{V}_{n(t,M)T/M}-V_{t}\big)^{2}\big]=\E\big[\check{V}_{n(t,M)T/M}-V_{t}\big]^{2}+\Var\big[\check{V}_{n(t,M)T/M}\big],
\end{equation*}
for proving the $L^{2}$-convergence rate it is sufficient to note that, in the same vein,
\begin{align*}
  \Var\big[\check{V}_{n(t,M)T/M}\big]&=\xi_{2}\sum^{n(t,M)-1}_{k=0}h^{2}\bigg(\frac{n(t,M)T}{M},\frac{kT}{M}\bigg)\frac{T}{M} +\varsigma^{2}\Var\Bigg[\cos\sum^{n(t,M)}_{k=1}\check{Z}_{k}\Bigg]\\
  &=\xi_{2}\int^{t}_{0}h^{2}(t,s)\dd s+\varsigma^{2}\Var[\cos Z_{t}]+O(M^{-1}),\quad\text{as }M\rightarrow\infty.
\end{align*}
$\square$

\subsection*{Proof of Proposition \ref{prop:4}}

The relationship between fractional moments and the characteristic function of a real-valued random variable has been well established. In particular, knowing that $I_{T}(\D)$ is strictly positive with $\E\big[I^{p}_{T}(\D)\big]<\infty$ for every $p>0$, we have ([Pinelis, 2016, Equation (2.19)] \cite{P2})
\begin{equation*}
  S^{(p)}_{t_{0}}=\E\big[\big(I^{2}_{T}(\D)\big)^{p/2}\big|\mathscr{F}_{t_{0}}\big]=(-\ii)^{p/2}\phi^{(p/2)}_{I^{2}_{T}(\D)|t_{0}}(0). \tag{A.5.1}
\end{equation*}
If $p$ is even, then (A.5.1) is understood as a conventional derivative corresponding to the first equation in (\ref{4.1.2}). Otherwise, it represents a fractional derivative and can be written ([Laue, 1980, Theorem 2.1] \cite{L1})
\begin{align*}
  S^{(p)}_{t_{0}}&=\sec\frac{\pi(p/2-\lfloor p/2\rfloor)}{2}\frac{p/2-\lfloor p/2\rfloor}{\Gf(1-p/2+\lfloor p/2\rfloor)} \\
  &\quad\times\Re\Bigg[(-\ii)^{\lfloor p/2\rfloor}\int^{\ell}_{-\infty}\frac{\phi^{(\lfloor p/2\rfloor)}_{I^{2}_{T}(\D)|t_{0}}(\ell)-\phi^{(\lfloor p/2\rfloor)}_{I^{2}_{T}(\D)|t_{0}}(l)}{(\ell-l)^{p/2-\lfloor p/2\rfloor+1}}\dd l\Bigg|_{\ell=0}\Bigg]. \tag{A.5.2}
\end{align*}
Since $\phi_{I^{2}_{T}(\D)|t_{0}}(\cdot)\in\mathcal{C}^{\lfloor p/2\rfloor}(\mathds{R})$, we can apply the dominated convergence theorem together with the substitution $l\mapsto-l$ to recast (A.5.2) as
\begin{align*}
  S^{(p)}_{t_{0}}&=\sec\frac{\pi(p/2-\lfloor p/2\rfloor)}{2}\frac{p/2-\lfloor p/2\rfloor}{\Gf(1-p/2+\lfloor p/2\rfloor)}\\
  &\quad\times\Re\Bigg[(-\ii)^{\lfloor p/2\rfloor}\int^{\infty}_{0}\frac{\phi^{(\lfloor p/2\rfloor)}_{I^{2}_{T}(\D)|t_{0}}(0)-\phi^{(\lfloor p/2\rfloor)}_{I^{2}_{T}(\D)|t_{0}}(-l)}{l^{p/2-\lfloor p/2\rfloor+1}}\dd l\Bigg].
\end{align*}
Using that $(-\ii)^{\lfloor p/2\rfloor}\phi^{(\lfloor p/2\rfloor)}_{I^{2}_{T}(\D)|t_{0}}(0)=S^{(2\lfloor p/2\rfloor)}_{t_{0}}$ and the Hermitian property of the characteristic function we arrive at the second equation in (\ref{4.1.2}).
$\square$

\subsection*{Proof of Corollary \ref{cor:1}}

Based on (\ref{2.3.14}), we have for $p/2\in\mathds{N}$ that
\begin{equation*}
  \triangle_{T}(\phi_{I^{2}_{T}(\D)|t_{0}}(l))=\ii l\phi_{I^{2}_{T}(\D)|t_{0}}(l),
\end{equation*}
so that
\begin{equation*}
  \big(\triangle_{T}(\phi_{I^{2}_{T}(\D)|t_{0}}(l))\big)^{(p/2)}=\ii\bigg(l\phi^{(p/2)}_{I^{2}_{T}(\D)|t_{0}}(l) +\frac{p\phi^{(p/2-1)}_{I^{2}_{T}(\D)|t_{0}}(l)}{2}\bigg).
\end{equation*}
Sending $l\rightarrow0$ gives the first equation in (\ref{4.1.4}).

For the second equation in (\ref{4.1.4}) we must justify that differentiation under $\triangle_{T}$ can be done inside the integral. Interchange with the real part is then simply allowed thanks to the Hermitian property. Since $\phi_{I^{2}_{T}(\D)|t_{0}}(\cdot)\in\mathcal{C}^{\lfloor p/2\rfloor}(\mathds{R})$ and $\phi^{(\lfloor p/2\rfloor)}_{I^{2}_{T}(\D)|t_{0}}(l)=O(\phi_{I^{2}_{T}(\D)|t_{0}}(l))$ as $l\rightarrow\infty$, we only need to check integrability of the tails of $\Re[\phi_{I^{2}_{T}(\D)|t_{0}}(l)]/l^{p/2-\lfloor p/2\rfloor}$ for $l\geq0$. From (A.2) we know that $J(t,t_{0},\D)>0$ and is measurable with respect to $\mathscr{F}_{t_{0}}$, allowing us to rewrite
\begin{equation*}
  \phi_{I^{2}_{T}(\D)|t_{0}}(l)=e^{\ii lJ(t,t_{0},X)}\varphi(l;T,t_{0},\D),\quad l\in\mathds{R},
\end{equation*}
where $\varphi(l;T,t_{0},\D)$ denotes the characteristic function of the random variable $\int^{T}_{t_{0}}H_{\D}(T,s)\dd X_{s}+\log\psi(-\ii,Z_{T}-Z_{t_{0}};t_{0},\D)>0$, whose distribution admits a well-defined density (recall that $H_{\D}$ is continuous, $\nu_{X}$ and $\nu_{Z}$ are both non-atomic and infinite). Hence, we have
\begin{equation*}
  \int^{\infty}_{0}\Re\big[e^{\ii lJ(t,t_{0},\D)}\varphi(l;T,t_{0},\D)\big]\dd l=0,
\end{equation*}
which with $p/2-\lfloor p/2\rfloor\in(0,1)$ for any $p\notin2\mathds{N}$ implies the desired tail integrability. $\square$

\subsection*{Proof of Proposition \ref{prop:5}}

Let $f(x;T,t_{0},\D)$ and $F(x;T,t_{0},\D)$, for $x>0$, respectively denote the density function and the distribution function of $I^{2}_{T}(\D)\big|\mathcal{F}_{t_{0}}$, which exist because the distributions of $X_{1}$ and $Z_{1}$ are both absolutely continuous. For the price of the asymmetric power put option on $I_{T}(\D)$ at $t_{0}\in[0,T)$, we adopt $\tilde{p}=p_{1}/2$ to rewrite its terminal payoff so that
\begin{align*}
  P^{(p_{1},p_{2},\rm(a))}_{t_{0}}&=\E\big[\big(K^{p_{2}}-I^{2\tilde{p}}_{T}(\D)\big)^{+}\big|\mathscr{F}_{t_{0}}\big]\\
  &=\int^{K^{p_{2}/\tilde{p}}}_{0}(K^{p_{2}}-x^{\tilde{p}})f(x;T,t_{0},\D)\dd x\\
  &=K^{p_{2}}F(K^{p_{2}/\tilde{p}};T,t_{0},\D)-\int^{K^{p_{2}/\tilde{p}}}_{0}x^{\tilde{p}}f(x;T,t_{0},\D)\dd x\\
  &:=\mathfrak{E}_{2}-\mathfrak{E}_{1}.
\end{align*}
Further denote $\tilde{K}=K^{p_{2}/\tilde{p}}$. Using the Fourier inversion formula we have
\begin{equation*}
  \mathfrak{E}_{2}=K^{p_{2}}\bigg(\frac{1}{2}-\frac{1}{\pi}\int^{\infty}_{0}\Re\bigg[\frac{e^{-\ii\tilde{K}l}\phi_{I^{2}_{T}(\D)|t_{0}}(l)}{\ii l}\bigg]\dd l\bigg)
\end{equation*}
and
\begin{equation*}
  \mathfrak{E}_{1}=\frac{1}{\pi}\int^{\tilde{K}}_{0}x^{\tilde{p}}\int^{\infty}_{0}\Re\big[e^{-\ii lx}\phi_{I^{2}_{T}(\D)|t_{0}}(l)\big]\dd l\dd x=\frac{1}{\pi}\int^{\infty}_{0}\Re\bigg[\phi_{I^{2}_{T}(\D)|t_{0}}(l)\int^{\tilde{K}}_{0}e^{-\ii lx}x^{\tilde{p}}\dd x\bigg]\dd l, \tag{A.7}
\end{equation*}
where the second equality uses the Fubini theorem since the integral in $x$ is taken over a finite interval. To evaluate the inner integral in (A.7), we apply the substitution $x\mapsto\ii lx$ and observe that
\begin{align*}
  \int^{\tilde{K}}_{0}e^{-\ii lx}x^{\tilde{p}}\dd x&=(\ii l)^{-\tilde{p}-1}\int^{\ii\tilde{K}l}_{0}e^{-x}x^{\tilde{p}}\dd x\\
  &=(\ii l)^{-\tilde{p}-1}\bigg(\int^{\infty}_{0}-\int^{\infty}_{\ii\tilde{K}l}\bigg)e^{-x}x^{\tilde{p}}\dd x\\
  &=(\ii l)^{-\tilde{p}-1}(\Gf(\tilde{p}+1)-\Gf(\tilde{p}+1,\ii\tilde{K}l)),
\end{align*}
where the second equality follows because the integrand is analytic over the horizontal half-strip $\{x:\Re x>0,\Im x\in(0,\tilde{K}l)\}$ with $l>0$. This establishes (\ref{4.2.3}) after rearrangement.

The pricing formula for the similar asymmetric call option results from a standard parity argument that
\begin{equation*}
  \big(I^{p_{1}}_{T}(\D)-K^{p_{2}}\big)^{+}-\big(K^{p_{2}}-I^{p_{1}}_{T}(\D)\big)^{+}=I^{p_{1}}_{T}(\D)-K^{p_{2}},
\end{equation*}
together with Proposition \ref{prop:4}. It is important to note that a single integral representation for the call price is inaccessible due to inapplicability of the Fubini theorem when integration acts over $[\tilde{K},\infty)\ni x$. $\square$

\subsection*{Proof of Corollary \ref{cor:2}}

We simply use the bounded-ness and Hermitian property of $\phi_{I^{2}_{T}(\D)|t_{0}}(\cdot)$ in order to apply $\triangle_{T}$ to (\ref{4.2.3}) inside the real part of the integral. For the call option we use the parity relation (\ref{4.2.4}). $\square$

\subsection*{Proof of Proposition \ref{prop:6}}

First consider the symmetric power put option with the payoff decomposition (\ref{4.3.2}), so that we may write
\begin{equation*}
  P^{(p,\rm(s))}_{t_{0}}=K^{p}F(K^{2};T,t_{0},\D)+\sum^{\infty}_{k=1}\binom{p}{k}(-1)^{k}K^{p-k}\bar{\mathfrak{E}}_{k}
\end{equation*}
and for every $k\in\mathds{N}_{++}$ using the argument in the proof of Proposition \ref{prop:5} we have
\begin{align*}
  \bar{\mathfrak{E}}_{k}&=\frac{1}{\pi}\int^{\infty}_{0}\Re\bigg[\phi_{I^{2}_{T}(\D)|t_{0}}(l)\int^{K^{2}}_{0}e^{-\ii lx}x^{k/2}\dd x\bigg]\dd l\\
  &=\frac{1}{\pi}\int^{\infty}_{0}\Re\bigg[\phi_{I^{2}_{T}(\D)|t_{0}}(l)\frac{\Gf(k/2+1)-\Gf(k/2+1,\ii K^{2}l)}{(\ii l)^{k/2+1}}\bigg]\dd l,
\end{align*}
which obviously allows the series to be augmented to $k=0$ and completes the proof of (\ref{4.3.4}) after simplification.

For the similar symmetric call option price, we rely on the decomposition (\ref{4.3.3}) to write
\begin{equation*}
  C^{(p,\rm(s))}_{t_{0}}=\sum^{\lfloor p\rfloor}_{k=0}\binom{p}{k}(-K)^{k}\breve{\mathfrak{E}}_{p-k}+\Sigma^{(p)}_{t_{0}},
\end{equation*}
where all the summands with index $k>p$ in conditional expectation are put into $\Sigma^{(p)}_{t_{0}}$. For every $0\leq k<p$ note that
\begin{equation*}
  \breve{\mathfrak{E}}_{p-k}=S^{(p-k)}_{t_{0}}-\int^{K^{2}}_{0}x^{(p-k)/2}f(x;T,t_{0},\D)\dd x,
\end{equation*}
so that a parity argument can be employed where the integral on the left-hand side is evaluated in the same vein as in (A.7). If $p\in\mathds{N}$ then $\Sigma^{(p)}_{t_{0}}$ is clearly naught. On the other hand, if $p\notin\mathds{N}$, then we write
\begin{equation*}
  \Sigma^{(p)}_{t_{0}}=\sum^{\infty}_{k=\lfloor p\rfloor+1}\binom{p}{k}(-K)^{k}\breve{\mathfrak{E}}_{k},
\end{equation*}
where
\begin{equation*}
  \breve{\mathfrak{E}}_{k}=\int^{\infty}_{K^{2}}x^{(p-k)/2}f(x;T,t_{0},\D)\dd x=\frac{1}{\pi}\int^{\infty}_{0}\Re\bigg[\phi_{I^{2}_{T}(\D)|t_{0}}(l)\int^{\infty}_{K^{2}}e^{-\ii lx}x^{-(k-p)/2}\dd x\bigg]\dd l.
\end{equation*}
Here the Fubini theorem applies because $k-p>0$. At this point it suffices to observe that
\begin{equation*}
  \int^{\infty}_{K^{2}}e^{-\ii l x}x^{-(k-p)/2}\dd x=\frac{\Gf(1-(k-p)/2,\ii K^{2}l)}{(\ii l)^{1-(k-p)/2}},
\end{equation*}
which is well-defined as $p-k$ cannot be an even number. $\square$

\subsection*{Proof of Corollary \ref{cor:3}}

The proof is similar to that of Corollary \ref{cor:2}, except that $\triangle_{T}$ acts on (\ref{4.3.4}) and (\ref{4.3.5}) termwise, where the interchange of integration and differentiation is permitted for the same reason. $\square$

\subsection*{Proof of Proposition \ref{prop:7}}

By mimicking the steps in the proof of Proposition \ref{prop:2}, it can be deduced from (\ref{6.2}) that the characteristic function of $\mathcal{T}_{t}$ for a fixed $t>0$ is given by
\begin{equation*}
  \phi_{\mathcal{T}_{t}}(l):=\E\big[e^{\ii l\mathcal{T}_{t}}\big]=\exp\int^{t}_{0}\log\phi_{Y_{1}}\bigg(l\int^{t}_{s}\eta(v,s)\dd v\bigg)\dd s,\quad l\in\mathds{R}.
\end{equation*}
By assumption the process $Y$ has its own filtration $\{\sigma((Y_{s})_{s\in[0,t]})\}_{t\geq0}$ independent from that of $X$. Therefore, via subsequent conditioning we have\footnote{To compute (A.11) one can also simulate $Y^{(\eta)}$, using what has been discussed in Section \ref{sec:3}.}
\begin{equation*}
  \phi_{\tilde{I}^{2}_{t}(\D)}(l):=\E\big[e^{\ii u\tilde{I}^{2}_{t}(\D)}\big]=\E\big[\E\big[e^{\ii uI^{2}_{\mathcal{T}_{t}}(\D)}\big|\sigma((Y_{s})_{s\in[0,t]})\big]\big]=\E\big[\phi_{I^{2}_{\mathcal{T}_{t}}(\D)|0}(l)\big]. \tag{A.11}
\end{equation*}
Since the distribution of the time change is absolutely continuous, (A.11) can be written using inverse Fourier transform as
\begin{equation*}
  \phi_{\tilde{I}^{2}_{t}(\D)}(l)=\frac{1}{\pi}\int^{\infty}_{0}\phi_{I^{2}_{s}(\D)|0}(l)\int^{\infty}_{0}\Re[e^{-\ii\ell s}\phi_{\mathcal{T}_{t}}(\ell)]\dd\ell\dd s,
\end{equation*}
and this is exactly the same as (\ref{6.3}). $\square$

\section*{Appendix B - A closed-form characteristic function}

In this section we prove a closed-form formula for the conditional characteristic function of $\tilde{V}^{\circ}_{t}(\D)$ given $\mathscr{F}_{t_{0}}$ for fixed $0\leq t_{0}<t\leq T$ and $\D\geq0$ using the tempered-stable distribution and the type-III kernel; note that $\tilde{V}^{\circ}_{t}(0)\equiv V^{\circ}_{t}$. The result will be useful for those who are interested in the pricing and hedging of derivatives contracts on the instantaneous or the forward variance under short-term dependence and volatility jumps, to which all the formulae presented in Section \ref{sec:4} apply. To that end let us recall that the partial forward variance $\tilde{V}^{\circ}_{t}(\D)$ has the general representation
\begin{equation*}
  \tilde{V}^{\circ}_{t}(\D)=\tilde{V}^{\circ}_{t_{0}}(t-t_{0}+\D)-\xi_{1}\int^{t}_{t_{0}}h(t+\D,s)\dd s+\int^{t}_{t_{0}}h(t+\D,s)\dd X_{s}. \tag{B.1}
\end{equation*}

\begin{corollary}
Let $h$ be the type-III kernel in its general form (\ref{2.2.11}), with $d\in(1/2,1)$, and let $X_{1}$ have the characteristic exponent (\ref{2.1.5}). Then we have
\begin{align*}
  \phi_{\tilde{V}^{\circ}_{t}(\D)|t_{0}}(l):=\E\big[e^{\ii l\tilde{V}^{\circ}_{t}(\D)}\big|\mathscr{F}_{t_{0}}\big]&=\exp\Bigg(\ii lI^{2}_{t_{0}}(t-t_{0}+\D) \\
  &\qquad+
  \begin{cases}
    \displaystyle \Psi_{-}(s)|^{t-t_{0}+\D}_{s=\D}\quad&\text{if }\tau>t-t_{0}+\D,\\
    \displaystyle \Psi_{-}(s)|^{\max\{\tau,\D\}}_{s=\D}+\Psi_{+}(s)|^{t-t_{0}+\D}_{s=\max\{\tau,\D\}}\quad&\text{if }\tau\leq t-t_{0}+\D
  \end{cases}
  \Bigg),\quad l\in\mathds{R}, \tag{B.2}
\end{align*}
where for $s\in[\D,t-t_{0}+\D]$
\begin{equation*}
  \Psi_{+}(s):=\frac{\ii la\theta\Gf(1-c)e^{-\kappa s}}{\kappa b^{1-c}}+a\Gf(-c)\bigg(\frac{e^{\kappa s}(b-\ii l\theta e^{-\kappa s})^{c+1}}{\ii lc\kappa\theta}\;_{2}\mathrm{F}_{1}\bigg(1,1;1-c;\frac{be^{\kappa s}}{\ii l\theta}\bigg)-b^{c}s\bigg) \tag{B.3}
\end{equation*}
and
\begin{align*}
  \Psi_{-}(s)&:=\frac{\ii la\Gf(1-c)s}{b^{1-c}}\bigg(\frac{d\tau^{d-1}-s^{d-1}}{\Gf(d+1)}-\theta e^{-\kappa\tau}\bigg)+a\Gf(-c)s\Bigg(\bigg(b-\ii l\theta e^{-\kappa\tau}+\frac{\ii l\tau^{d-1}}{\Gf(d)}\bigg)^{c} \\
  &\qquad\times\;_{2}\mathrm{F}_{1}\bigg(-c,\frac{1}{d-1};\frac{d}{d-1};\frac{\ii ls^{d-1}}{\ii l\tau^{d-1}+(b-\ii l\theta e^{-\kappa\tau})\Gf(d)}\bigg)-b^{c}\Bigg). \tag{B.4}
\end{align*}
\end{corollary}

\textsl{Proof.} Following the proof of Proposition \ref{prop:2} it is readily established from (B.1) that
\begin{equation*}
  \phi_{\tilde{V}_{t}(\D)|t_{0}}(l)=\exp\bigg(\ii l\bigg(\tilde{V}^{\circ}_{t_{0}}(t-t_{0}+\D)-\xi\int^{t}_{t_{0}}h(t+\D,s)\dd s\bigg)+\int^{t}_{t_{0}}\log\phi_{X_{1}}(lh(t+\D,s))\dd s\bigg). \tag{B.5}
\end{equation*}
Rewriting the type-III kernel as
\begin{equation*}
  h(t-s+\D)=
  \begin{cases}
    h_{-}(t-s+\D),&\quad\text{if }t-s+\D<\tau,\\
    h_{+}(t-s+\D),&\quad\text{if }t-s+\D\geq\tau,
  \end{cases}
\end{equation*}
with
\begin{equation*}
  h_{-}(t-s+\D):=\frac{(t-s+\D)^{d-1}-\tau^{d-1}}{\Gf(d)}+\theta e^{-\kappa\tau}\quad\text{and}\quad h_{+}(t-s+\D)=\theta e^{-\kappa(t-s+\D)},
\end{equation*}
straightforward integration over the interval $[t_{0},t]$ thus leads to
\begin{align*}
  &\quad\int^{t}_{t_{0}}h(t-s+\D)\dd s\\
  &=
  \begin{cases}
    \displaystyle \int^{t}_{t_{0}}h_{-}(t-s+\D)\dd s&\quad\text{if }\tau>t-t_{0}+\D,\\
    \displaystyle \int^{t}_{\min\{t+\D-\tau,t\}}h_{-}(t-s+\D)\dd s+\int^{\min\{t+\D-\tau,t\}}_{t_{0}}h_{+}(t-s+\D)\dd s&\quad\text{if }\tau\leq t-t_{0}+\D
  \end{cases}\\
  &=
  \begin{cases}
    \displaystyle \int^{t-t_{0}+\D}_{\D}h_{-}(s)\dd s&\quad\text{if }\tau>t-t_{0}+\D,\\
    \displaystyle \int^{\max\{\tau,\D\}}_{\D}h_{-}(s)\dd s+\int^{t-t_{0}+\D}_{\max\{\tau,\D\}}h_{+}(s)\dd s&\quad\text{if }\tau\leq t-t_{0}+\D
  \end{cases}\\
  &=
  \begin{cases}
    \displaystyle s\bigg(\frac{s^{d-1}-d\tau^{d-1}}{\Gf(d+1)}+\theta e^{-\kappa\tau}\bigg)\bigg|^{t-t_{0}+\D}_{s=\D}&\quad\text{if }\tau>t-t_{0}+\D,\\
    \displaystyle s\bigg(\frac{s^{d-1}-d\tau^{d-1}}{\Gf(d+1)}+\theta e^{-\kappa\tau}\bigg)\bigg|^{\max\{\tau,\D\}}_{s=\D}-\frac{\theta e^{-\kappa s}}{\kappa}\bigg|^{t-t_{0}+\D}_{s=\max\{\tau,\D\}}&\quad\text{if }\tau\leq t-t_{0}+\D,
  \end{cases}
\end{align*}
where the second equality uses the substitution $s\mapsto t-s+\D$ and $\int^{\max\{\tau,\D\}}_{\D}\equiv0$ if $\D\geq\tau$.

Similarly, for the second Riemann integral in (B.5) we have with the characteristic exponent (\ref{2.1.5}) that $\xi_{1}=a\Gf(1-c)/b^{1-c}$ and that
\begin{align*}
  &\quad\int^{t}_{t_{0}}\log\phi_{X_{1}}(lh(t-s+\D))\dd s\\
  &=
  \begin{cases}
    \displaystyle \int^{t-t_{0}+\D}_{\D}\log\phi_{X_{1}}(lh_{-}(s))\dd s&\quad\text{if }\tau>t-t_{0}+\D,\\
    \displaystyle \int^{\max\{\tau,\D\}}_{\D}\log\phi_{X_{1}}(lh_{-}(s))\dd s+\int^{t-t_{0}+\D}_{\max\{\tau,\D\}}\log\phi_{X_{1}}(lh_{+}(s))\dd s&\quad\text{if }\tau\leq t-t_{0}+\D.
  \end{cases} \tag{B.6}
\end{align*}
Since the integrands in (B.6) are obviously integrable over the designated domains, it only suffices to consider the indefinite integrals, $\mathcal{I}_{+}(s):=\int(b-\ii le^{-\kappa s})^{c}\dd s$ and $\mathcal{I}_{-}(s):=\int(b-\ii ls^{d-1})^{c}\dd s$. Note that $a\Gf(-c)$ is just a scaling factor while the integration of $b^{c}$ is immediate. For $\mathcal{I}_{+}$, we observe by using binomial expansion that
\begin{align*}
  \mathcal{I}_{+}(s)&=(-1)^{c+1}(\ii l)^{c}\sum^{\infty}_{k=0}\binom{c}{k}\bigg(-\frac{b}{\ii l}\bigg)^{k}\frac{e^{-\kappa(c-k)s}}{\kappa(c-k)}\\
  &=\frac{(-1)^{c+1}(\ii l)^{c}e^{-\kappa cs}}{c\kappa}\sum^{\infty}_{k=0}\frac{(-c)^{2}_{k}}{(1-c)_{k}}\bigg(\frac{be^{\kappa ks}}{\ii l}\bigg)^{k}\\
  &=\frac{(-1)^{c+1}(\ii l)^{c}e^{-\kappa cs}}{c\kappa}\;_{2}\mathrm{F}_{1}\bigg(-c,-c;1-c;\frac{be^{\kappa ks}}{\ii l}\bigg),
\end{align*}
where $(\cdot)_{\cdot}$ denotes the Pochhammer symbol, \text{a.k.a.} the rising factorial, and which after simplification leads to (B.3). The case of $\mathcal{I}_{2}$ is slightly more involved but can be proved in a similar fashion, and we obtain
\begin{equation*}
  \mathcal{I}_{-}(s):=s\Bigg(\bigg(b-\ii le^{-\kappa\tau}+\frac{\ii l\tau^{d-1}}{\Gf(d)}\bigg)^{c}\;_{2}\mathrm{F}_{1}\bigg(-c,\frac{1}{d-1};\frac{d}{d-1};\frac{\ii ls^{d-1}}{\ii l\tau^{d-1}+(b-\ii le^{-\kappa\tau})\Gf(d)}\bigg)\Bigg),
\end{equation*}
which eventually yields (B.4). Putting things together we arrive at (B.2) as required. $\square$

We remark that, with $X_{1}$ being tempered-stable distributed, if $h$ is a kernel of either type I or type II, no such explicit expression exists for the second Riemann integral in (B.5), the computation of which has to resort to numerical methods such as the Gauss quadrature rule. The reason behind this problem is clear: the type-I and type-II kernels are both formed by multiplying power and exponential functions but no elementary substitution works for integrals of the form $\int(1-e^{-\kappa s}s^{d-1})^{c}\dd s$, for $d>1/2$, $\kappa>0$ and $c\in(0,1)$.

\end{document}